\documentclass[aps,prc,amsfonts,preprintnumbers,superscriptaddress,showpacs,nofootinbib]{revtex4}
\pdfoutput=1
\usepackage{graphics}
\usepackage{amssymb}
\usepackage{psfrag}
\usepackage{epsfig}
\usepackage{epsf}
\usepackage{float}
\usepackage{amsmath,amsthm}
\allowdisplaybreaks

\usepackage{xparse}
\ExplSyntaxOn
\NewDocumentCommand{\mref}{m}{\quinn_mref:n {#1}}
\seq_new:N \l_quinn_mref_seq
\cs_new:Npn \quinn_mref:n #1
 {
  \seq_set_split:Nnn \l_quinn_mref_seq { , } { #1 }
  \seq_pop_right:NN \l_quinn_mref_seq \l_tmpa_tl
  ( 
  \seq_map_inline:Nn \l_quinn_mref_seq
    { \ref{##1},\nobreakspace } 
  \exp_args:NV \ref \l_tmpa_tl 
  ) 
 }
\ExplSyntaxOff

\newcommand{\fet}[1]{\mbox{\boldmath $#1$}}

\newcommand{\nn}{\nonumber \\ }

\begin{document}

\title{Low-energy theorems for nucleon-nucleon scattering at  $M_{\pi}=450$ MeV}

\author{V.~Baru}
\affiliation{Institut f\"ur Theoretische Physik II, Ruhr-Universit\"at Bochum,
  D-44780 Bochum, Germany}
\affiliation{Institute for Theoretical and Experimental Physics, B. Cheremushkinskaya 25, 117218 Moscow, Russia}

\author{E.~Epelbaum}
\affiliation{Institut f\"ur Theoretische Physik II, Ruhr-Universit\"at Bochum,
  D-44780 Bochum, Germany}

\author{A.~A.~Filin}
\affiliation{Institut f\"ur Theoretische Physik II, Ruhr-Universit\"at Bochum,
  D-44780 Bochum, Germany}

\date{\today}

\begin{abstract}
We apply the low-energy theorems to analyze the recent lattice QCD
results for the two-nucleon system at a pion mass of $M_\pi\simeq
450$~MeV obtained by the NPLQCD collaboration.
We find that the binding energies of the deuteron and dineutron
are  inconsistent  with the low-energy behavior of the corresponding
phase shifts within the quoted uncertainties and vice versa.
Using the binding energies of the deuteron and dineutron as
input, we employ the low-energy theorems to predict the phase shifts
and extract the scattering length and
the effective range in the  $^3S_1$  and $^1S_0$ channels.  Our
results for these quantities are consistent with those obtained by
the NPLQCD collaboration from effective field theory analyses but are in conflict with their
determination based on the effective-range approximation.
 \end{abstract}


\maketitle

\vspace{-0.2cm}

\section{Introduction}
\def\theequation{\arabic{section}.\arabic{equation}}
\label{sec:intro}

Understanding of certain fine-tunings in the parameters of the
Standard Model is an important frontier in modern hadron and
nuclear physics. In connection with the anthropic considerations, a
question has been raised whether
the light quark masses have to take very specific values in order
to maintain conditions essential for the development of life, see
Ref.~\cite{Meissner:2014pma} for discussion.
In particular,  the proximity  of  the Hoyle state, the first $0^+$ excited  state
of $^{12}$C, to  the  triple alpha-particle threshold is known to be
crucial for the enhanced resonance  formation of  the life-important
elements  $^{12}$C  and  $^{16}$O  in red giant stars~\cite{Oberhummer:2000mn}.
The dependence of the excitation energy  of the Hoyle state on the
light-quark masses was analyzed recently within {\it ab initio}
nuclear lattice simulations~\cite{Epelbaum:2012iu,Epelbaum:2013wla}.
It was found that the variation of the light quark masses by a few
percent  is likely to be not detrimental for the development of
life. More conclusive statements would require a better knowledge of
the quark mass
dependence of the nuclear force  or, more precisely, of the
nucleon-nucleon (NN) $S$-wave scattering lengths which is 
by far the dominant source of theoretical uncertainty in this
calculation.
The quark mass dependence of the nuclear force also plays an important role
for constraining a time variation of the Standard Model
parameters as predicted by various extensions of the Standard Model at
the time of Big Bang nucleosynthesis by comparing the
observed and calculated primordial deuterium and helium abundances
\cite{Bedaque:2010hr,Berengut:2013nh}.

In  recent years, there has been significant progress in
lattice-QCD calculations of nuclear systems which constitute  the
primary  source  of information about
the light-quark or, equivalently, pion mass dependence of
nuclear observables.   In particular,  the fully dynamical
calculations at unphysical pion masses as low as $M_\pi\simeq 300$-$400$~MeV
have been performed, see
e.g.~Refs.~\cite{Beane:2011iw,Yamazaki:2015asa,Orginos:2015aya}.
To connect these results with experimentally observed quantities corresponding
to the physical value of the quark masses one can employ chiral
effective field theory (EFT) which is still
expected to be applicable at such pion masses 
\cite{Beane:2001bc,Beane:2002vs,Epelbaum:2002gb,Beane:2002xf,Epelbaum:2002gk,Chen:2010yt,Soto:2011tb,Epelbaum:2013ij,Berengut:2013nh}.
Notice that not only the binding energies but
also the NN scattering observables such as the  phase shifts and effective
range parameters have been calculated on a lattice,  see
Refs.~\cite{Orginos:2015aya,Beane:2012vq,Beane:2013br} for recent analyses of the
NN
$^1S_0$  and $^3S_1$ channels  by the NPLQCD  collaboration at $M_\pi
\simeq 450$ and $800$ MeV and  Ref.~\cite{Berkowitz:2015eaa}  for  the
analysis of the higher partial waves  at   $M_\pi \simeq 800$ MeV by
the CalLat collaboration.
Interestingly, the general trend of  lattice calculations by different
groups suggests a stronger  attraction
both in the $^1S_0$  and $^3S_1$ channels  when going  away  from the physical point
towards heavier quark masses
\cite{Beane:2011iw,Yamazaki:2015asa,Orginos:2015aya,Beane:2013br,Yamazaki:2012hi}.
Both the  deuteron and  dineutron systems in
these calculations are  found to be bound at unphysically heavy pion masses with  the
deuteron binding energy being significantly larger than the experimentally  observed one.    These
results  are, however,  not supported by the HAL QCD collaboration
which  finds  no bound states  in  these channels  for pion masses
ranging from
$469$ to $1171$ MeV~\cite{Inoue:2011ai}.  It should be noted that
unlike the lattice calculations mentioned above,  a different approach
 is employed by the HAL QCD collaboration, which makes use  
of a two-nucleon  potential at the intermediate step.
 The  puzzle is  getting even more intriguing  given that the chiral
 EFT  calculations tend to indicate less attraction  at
 $M_\pi$ larger than the physical value~\cite{Beane:2001bc,Beane:2002xf,Epelbaum:2002gb,Berengut:2013nh,
 Epelbaum:2013ij}, see also Ref.~\cite{Flambaum:2007mj} for a related
work.  These  calculations, however, rely on the naturalness assumption
and/or make use of resonance saturation estimates for 
$M_\pi$-dependent  four nucleon
contact interactions.

\begin{figure}[tb]
\vspace*{-0.cm}
\includegraphics[width=0.35\textwidth,keepaspectratio,angle=0,clip]{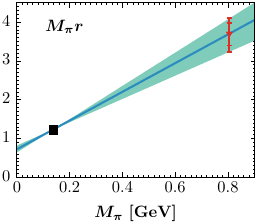}
    \caption{
\label{fig:3s1eff_rad_mpi} (Color online) Linear with $M_\pi$ behavior  of
      $M_\pi r$  in  the $^3$S$_1$     partial wave conjectured in
      Ref.~\cite{Beane:2013br}.  Red solid triangles correspond to the
      NPLQCD results at $M_\pi \simeq 800$~MeV \cite{Beane:2012vq} while the shaded
      area shows the uncertainty of the suggested linear
      interpolation.  The black squares show 
      the empirical values of the effective range at the
      physical pion mass \cite{deSwart:1995ui}.
}
\end{figure}

In our recent paper~\cite{Baru:2015ira}, we have argued that the
low-energy theorems (LETs) in NN scattering can provide important
consistency
checks of both lattice-QCD results and their
chiral extrapolations.    Specifically,  the knowledge of the analytic
properties of the scattering amplitude allows one to predict its
energy dependence and, under certain circumstances,
extract the parameters of  the effective range expansion.
In Ref.~\cite{Baru:2015ira},  we have tested this
approach by predicting the  effective range parameters for $S$-wave NN
scattering in  the spin-triplet and the spin-singlet channels, see
also
Refs.~\cite{Cohen:1998jr,Cohen:1999iaa,Steele:1998zc,Epelbaum:2009sd,Epelbaum:2009zz,Epelbaum:2010nr}
for  earlier  studies along this line.
Further,  we generalized the  LETs  to unphysical pion masses and
applied the resulting approach to selected lattice-QCD results. 
In particular,  using  the  linear behavior of  the
quantity $M_\pi r$, where $r$ refers to the effective range, conjectured in
Ref.~\cite{Beane:2013br},  and visualized in Fig.~\ref{fig:3s1eff_rad_mpi},
we employed  the LETs  to predict the
$M_\pi$  dependence of the deuteron binding energy  and of the other
parameters  in the effective range expansion of the NN scattering
amplitude in the $^3S_1$ channel.  Remarkably,  the resulting 
$M_\pi$-dependence of the deuteron binding energy  turned out to be in good agreement with the
general trend  of  different lattice  calculations
\cite{Beane:2011iw,Yamazaki:2015asa,Orginos:2015aya,Beane:2013br,Yamazaki:2012hi}
except for the results of
Ref.~\cite{Inoue:2011ai}  which do not support the existence of
bound states for large pion masses.

 Recently,  the NPLQCD collaboration has reported their new results
 for  the $S$-wave NN observables both in the spin-triplet and the
 spin-singlet channels at  $M_\pi=450$ MeV~\cite{Orginos:2015aya}.
 In particular,  they have calculated the values of the deuteron
 and the dineutron binding energies and the phase
 shifts  in the $^3S_1$
 and $^1S_0$ channels  at several values of the center of mass (cms) momenta
 above threshold.  In this paper we confront these
 results with the LETs.
In particular, we demonstrate that the lattice phase shifts at the two
lowest energies in the
$^3S_1$ channel are inconsistent (within the quoted uncertainties) 
with  the  deuteron binding energy
obtained in the same lattice calculation.  The situation is less
conclusive in the $^1S_0$ channel due to larger uncertainties of the
LETs in this channel.  However,  the results of our analysis in this channel
also indicate an inconsistency between the phase shifts and the large
value of the dineutron binding energy reported by the NPLQCD collaboration.
Taking the NPLQCD results for the  deuteron and
dineutron binding energies as input, we use the LETs to infer the corresponding
phase shifts and extract the values of the scattering lengths and
effective ranges. We perform a detailed comparison of our results for
these quantities with the ones of Ref.~\cite{Orginos:2015aya} and argue that their
determination by means of the effective range approximation is
not self-consistent. 

Our paper is organized as follows. In section~\ref{sec:LETphys}, we
discuss in detail our formalism, explain the
meaning of the LETs and discuss their generalization   to unphysical values
of the pion mass.
Implications of the LETs for the recent lattice-QCD results at $M_\pi \sim
450$~MeV are considered in section~\ref{sec:LatticeQCD}.
Section~\ref{sec:rangempi} addresses the dependence of the effective range on the pion mass.
Finally, our main findings are summarized in section~\ref{sec:Summary}.

\section{Low-energy theorems for NN scattering}
\def\theequation{\arabic{section}.\arabic{equation}}
\label{sec:LETphys}

\subsection{The formalism}

The  concept of the low-energy theorems  for NN scattering and their
generalization to unphysical pion masses have been discussed
in Ref.~\cite{Baru:2015ira},
see also
Refs.~\cite{Cohen:1998jr,Cohen:1999iaa,Steele:1998zc,Epelbaum:2009sd,Epelbaum:2009zz,Epelbaum:2010nr}
for  related earlier studies.
In this section we formulate the main idea of the LETs from a somewhat
different perspective as compared to Ref.~\cite{Baru:2015ira},
where a quantum mechanical framework of the modified effective range
expansion~\cite{vanHaeringen:1981pb} was employed.

We assume that the NN interaction is characterized  by two distinct
scales $ M_L$ and   $M_S$, $ M_L \ll M_S$, so that the
potential can be written as
\begin{equation}
V=V_L  +V_S\,,
\end{equation}
with the interaction ranges of the order of $r_L \sim M_L^{-1}$  and  $r_S \sim M_S^{-1} $, respectively.
The analytic structure of  the scattering amplitude near threshold is
governed by the long-range  interactions.
Taking into account the discontinuity across the left-hand cut from
the  long-range potential $V_L$, the energy dependence of the scattering amplitude
can be predicted in a model independent way
up to  the energies  corresponding to
the branch point of  a more distant  left-hand cut  associated with
the potential $V_S$.
This prediction can be regarded as  a low-energy theorem.  Alternatively,  one
can view  the LETs as  correlations between the parameters in the
effective-range expansion of  the inverse scattering amplitude induced by
long-range interactions.  Notice that 
the inverse scattering amplitude may possess poles in the near-threshold region,
whose appearance 
does not  affect the  validity range of the LETs if
the scattering amplitude  is  kept  unexpanded.

The longest-range part of the NN  force is  due to the  one-pion
exchange potential (OPEP). Thus, the LET for NN scattering are
expected to be governed by the left-hand cut generated by  the OPEP.
The  OPEP  is, however, singular  at the origin and  requires
regularization and  renormalization.
Therefore, instead of using the quantum mechanical approach,
we  formulate  the  LET  within  the  modified  Weinberg approach of  chiral EFT
\cite{Baru:2015ira,Epelbaum:2012ua,Epelbaum:2013ij,Epelbaum:2015sha}.
Since  the correlations  between  the effective range expansion parameters   are inherently
long-range  phenomena,  the results after renormalization and removing
the ultraviolet cutoff should be model and regularization-scheme independent.

To be specific, we calculate the  
 scattering  amplitude  $T$ by solving
the Lippmann-Schwinger-type integral  equation introduced originally by
 Kadyshevsky~\cite{Kadyshevsky:1967rs} which, for the case of the
 fully off-shell kinematics, has the form
\begin{equation}
T\left(
\vec p, \vec p\,' , k\, \right)=V \left(
\vec p \, , \vec p\,'\right) + \int d^3 q \;
 V  \big(
\vec p, \vec q \, \big) \ G(k, q) \ T  \big(  \vec q,\vec p\,', k \big)
\,,\label{MeqLOk0integrated}
\end{equation}
where $G(k ,q)$ is the free Green function
\begin{equation}
G(k, q)=\frac{m_N^2}{2(2\,\pi)^3}\frac{1}{\big(\vec q\:^2+m_N^2\big)\left(
E_k-\sqrt{\vec q\:^2+m_N^2}+i \epsilon\right)}\,.
\label{G}
\end{equation}
Further, $\vec p$ ($\vec p \, '$) is the incoming (outgoing) three-momentum of
the nucleon in the cms and $E_k = \sqrt{\vec k  ^2 +
  m_N^2}$ with $m_N$ denoting the nucleon mass and $\vec k$ being the
corresponding (on-mass-shell)  three-momentum.
The $S$-wave potential  at leading order (LO)  consists  of the OPEP  and two
derivative-less  contact interactions (here denoted $C_0$  for each given partial wave)
\begin{equation}
V_{\rm LO}\big(
\vec p, \vec p\,' \, \big)=    - \frac{g_A^2}{4 F_\pi^2} \, \frac{\vec \sigma_1 \cdot 
(\vec p- \vec p\,') \; \vec \sigma_2 \cdot (\vec p- \vec p\,')}{(\vec p- \vec p\,') \, ^2 + M_\pi^2} \,
{ \fet\tau_1 \cdot  \fet\tau_2}  +C_0\, ,
\label{VLO}
\end{equation}
where $\vec \sigma_i$ ($\fet \tau_i$) denote the spin
(isospin) Pauli matrices of the nucleon $i$,  while  $g_A$ and
$F_\pi$ refer to the axial vector coupling of the nucleon and the pion
decay constant, respectively.
As  discussed in Ref.~\cite{Epelbaum:2012ua},   the  LO  integral
equation in this  framework is exactly renormalizable\footnote{For a
  recent extension of the approach to $\bar D D^*$ scattering see
  Ref.~\cite{Baru:2015tfa}. This paper
  also addresses chiral extrapolations of the
  $X(3872)$ binding energy.},  that is  all
ultraviolet divergencies  appearing from iterations
of the LO potential can be removed via an appropriate redefinition  of the  contact
interaction $C_0$.  As a  consequence,  the cutoff  in the  integral
equation  can be put  to infinity,  so that  no finite-cutoff
artifacts  can   affect  the  LET.

\begin{figure}[tb]
\vskip 1 true cm
\includegraphics[width=0.9\textwidth,keepaspectratio,angle=0,clip]{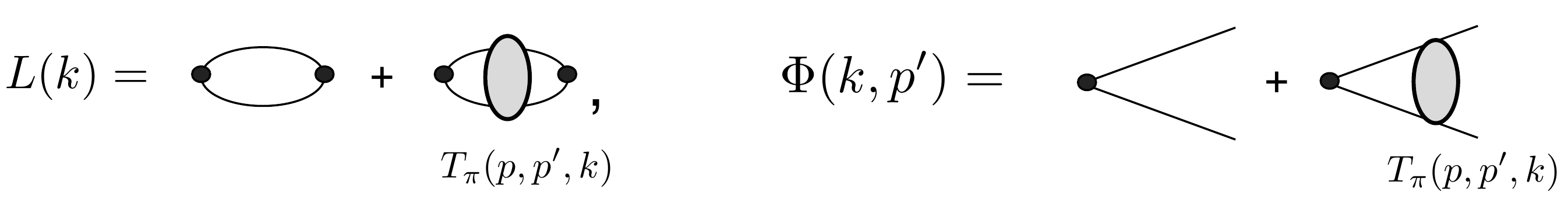} 
    \caption{
\label{fig:blocks}
Graphical illustration  of the functions $\Phi$  and $L$ in
Eq.~\eqref{fun}, which correspond to convolutions of the  pionic
amplitude $T_{\pi}$ with the point-like vertices as explained in the text.}
\end{figure}

The integral equation (\ref{MeqLOk0integrated}) with the potential
(\ref{VLO}) can,  in general,  be only  solved numerically.
However, the correlations between the parameters of the effective
range expansion implied by the LETs can be demonstrated analytically.
For simplicity reasons,   we consider the partial wave projected $T$-matrix in an  uncoupled  channel  with the zero orbital angular 
momentum,  whereas  generalizations to the 
coupled-channel case and to non-zero angular momenta are straightforward.
Since the LO contact interaction is separable, it is possible to write the solution for the $T$-matrix in the semi-analytic (operator) form
\begin{equation}
\label{separ}
T( p, p^\prime ,k) = T_{\pi}( p, p^\prime ,k) + \Phi(
p,k)\,  D(k)\,  \Phi(k, p^\prime ), \quad \quad
D(k) = \frac1{C_0^{-1} - L(k)},
\end{equation}
where  
 $T_{\pi}( p, p^\prime ,k) $  denotes the  off-shell
solution of  the projected integral equation (\ref{G})  with the OPEP alone ($C_0=0$),
whereas the functions $\Phi$  and  $L$,  shown graphically
in Fig.~\ref{fig:blocks}, involve convolutions of this pionic
amplitude with the  point-like  vertex,  namely
\begin{eqnarray}
\label{fun}
\nonumber
\Phi(k,  p^\prime ) &=&   1+\int d q  \ q^2\,  G(k, q) \ T_{\pi}  \big( q, p^\prime, k \big)  ,\\
\Phi( p, k) &=&   1+\int  d q  \ q^2\,  T_{\pi}  \big(   p,
q, k \big)\  G(k, q), \\
L(k ) &=&  
\int d q  \ q^2\,  G(k, q) 
+ 
\int   d q \, d q^\prime \, q^2  \, q^{\prime2} \, G(k, q) \, T_{\pi}  \big(   q, q^\prime, k \big) \,  G(k, q^\prime) \,.
\nonumber
\end{eqnarray}

To complete the renormalization program at LO, one has to express the
contact interaction $C_0$  in terms of the scattering amplitude at zero
momentum, that is the scattering length $a$,
which is assumed to be an input quantity.
Thus,  the on-shell NN amplitude reads
\begin{equation}
\label{tmaton}
T(k) = T_{\pi}(k) + \Phi(k)\,  D(k)\,  \Phi(k), \quad \quad
  D(k) = \frac1{ \frac{\Phi^2(0)}{T(0)-T_{\pi}(0)} +L(0) - L(k)},
  \quad  \quad  T(0) = -\frac{16 \pi^2}{m_N} a\,,
\end{equation}
where  we used the shorthand notation  $T(k) \equiv T(k,k,k);  \
\Phi(k) \equiv \Phi(k,k) $.

For an uncoupled channel with the zero orbital angular momentum,
the scattering amplitude $T(k)$   can  be
expressed in terms of the so-called effective range function $F (k)
\equiv k\, {\rm cot}\, \delta (k)$ via
\begin{equation}
\label{Fk}
T (k) = -\frac{16 \pi^2}{m_N} \frac{1}
{F (k) - i k} \,.
\end{equation}
We are now in the position to  formulate the LETs at LO: Using the
scattering length as the only input quantity to fix the unknown
low-energy constant, we employ
Eqs.~(\ref{tmaton})-(\ref{Fk}) to predict the momentum dependence of
the scattering amplitude and to calculate the phase shifts $\delta (k)$.   Such a
prediction is possible because all information about the longest-range OPE
potential and, in particular, about the discontinuity across the
corresponding left-hand cut starting from  the
momentum   $k = \pm i M_{\pi}/2$,  is explicitly taken into account in the
calculation.

Unlike  the scattering amplitude, the effective range function does
not possess the kinematic
unitarity cut and is a  real meromorphic function of $k^2$ near the origin $k=0$
\cite{Blatt:49,Bethe:49}.
It can, therefore, be Taylor-expanded about the origin leading to the   effective
range expansion\footnote{We assume here that the phase shift does not cross
  zero in the region of validity of the effective range expansion. If this is the case, the
  Taylor expansion should be replaced by
e.g.~the Pad\'{e} approximation or  the
amplitude should be kept unexpanded, see Ref.~\cite{Midya:2015eta} for
a related discussion.
}
\begin{equation}
\label{ere}
 k \,{\rm cot}\, \delta (k) = - \frac{1}{a} + \frac{1}{2} r k^2 +
 v_2 k^4 +  v_3 k^6 +  v_4 k^8 + \ldots\,,
\end{equation}
where   $r$  is the effective
range,  while $v_i$ are the so-called shape parameters.
Thus, a single  piece of information in the form of the scattering
length  (or the energy of a bound/virtual state)  allows one to
predict all the coefficients in  the  effective range expansion
\begin{equation}
\! \! \! r= \frac{\alpha (a M_{\pi})}{M_{\pi}},  \quad \quad   v_i=
\frac{\beta_i (a M_{\pi})}{M_{\pi}^{2i-1}},
\end{equation}
where  $\alpha$ and  $\beta_i$ are polynomials in the inverse
scattering length,  namely
\begin{eqnarray}
\alpha&=& \alpha_0 + \frac{\alpha_1}{(a M_\pi )}+
\frac{\alpha_2}{(aM_\pi )^2 }\,, \\
\beta_i&=& \beta_{i,  0} + \frac{\beta_{i,  1}}{(a M_\pi )}+
\frac{\beta_{i,  2}}{(aM_\pi )^2 } + \ldots + \frac{\beta_{i,  i+1}}{(aM_\pi )^{i+1} }\,,
\end{eqnarray}
with the coefficients $\alpha_i$ and $\beta_{i,j}$ being calculable
from the various quantities appearing in Eq.~(\ref{tmaton}) and their
derivatives evaluated at $k=0$.  Their explicit form can be easily
obtained by performing Taylor expansion of the inverse amplitude
$T^{-1}$ around $k^2 = 0$.

Given that  the  left-hand cut  from the OPEP  is explicitly included, the convergence
radius of the LETs is restricted by the next-to-lowest-lying left-hand
singularity associated with the two-pion exchange potential (TPEP). 
In the present analysis, we take into account the contributions of the
TPEP implicitly by  including the relevant  momentum-dependent
short-range interactions  at next-to-leading order (NLO),
whose strengths have to be adjusted to
reproduce the empirical value of the effective range in the
corresponding channel.  Then,  the  shape
parameters  can  be predicted  up-to-and-including  the NLO
corrections.

After renormalization, the predictions for  the LETs become
insensitive to details of the short-range interaction once its
strength is adjusted to reproduce the physical observable.
As argued in Ref.~\cite{Baru:2015ira}, it is convenient to employ
resonance saturation via a heavy-meson exchange in order to model
higher-order contact interaction without destroying explicit renormalizability
of the integral equation or having to rely on perturbation theory.
Specifically, the NLO correction to the potential is taken in the form
\begin{equation}
\label{satur}
V_{\rm NLO} \big(
\vec p, \vec p\,' \, \big)= \beta \,\frac{\vec \sigma_1 \cdot 
(\vec p- \vec p\,') \; \vec \sigma_2 \cdot (\vec p- \vec p\,')}{(\vec p- \vec p\,') \, ^2 + M^2} ,  
\end{equation}
where the heavy-meson mass $M$ is set to be $M=700$~MeV and the strength
$\beta$ is adjusted to reproduce the empirical value of the effective
range in the $^1S_0$ and the $^3S_1$ channels. 
Our results are not sensitive to the functional form of the term
parameterizing the subleading short-range interaction in
Eq.~(\ref{satur}), see Ref.~\cite{Baru:2015ira} for more details.

We now summarize the main findings of Ref.~\cite{Baru:2015ira} for the
physical value of the pion mass:
\begin{itemize}
\item
In the $^3S_1$ channel, the LETs  yield  very accurate results
already at  LO. For example,  the effective range is predicted with an
accuracy  better than $10\%$.
The   accuracy  at LO  
even appears to be better than one could naively expect from the ratio of scales
corresponding to the explicitly included lowest left-hand cut
from the OPEP  and the next-to-lowest one from the TPEP, which is not
considered explicitly  at  this order.
This observation can be understood by noticing that iterations of the
OPEP do actually generate the dominant contributions to the left-hand
cuts due to the two- and multiple-pion exchange. 
\item
The accuracy of the LETs  in the $^1S_0$ channel is much worse
  than in the spin-triplet case. This has to be expected due to the
  weakness of the OPEP in that channel.
  Indeed,  the OPEP  contributes less than $20\%$  to the  magnitude of
  the  empirical  $^1S_0$  phase  shift at its maximum value.
\item
As expected, the predicted values of the
shape parameters at  NLO show in both channels a clear improvement as compared with the
LO results. In particular, the NLO LETs appear to be accurate at the level of a few
percent for the $^3S_1$ channel  (except for $v_2$ which is unnaturally
small),  while  the  accuracy of  the predictions in the $^1S_0$
channel is  improved to $\sim 25\%$.
\end{itemize}

\subsection{Generalization to unphysical pion masses}
\label{general}

A generalization of the LETs to the case of unphysical pion masses can
be carried out straightforwardly~\cite{Baru:2015ira}.
The main dynamical effect of changing the pion mass in the OPEP
corresponds to shifts of  the branch points of  all left-hand cuts.
The discontinuity across the  left-hand cuts also changes due to the
dependence of the ratio
$g_A/F_\pi$, which determines the strength of the OPEP, on the pion
mass\footnote{We do not take into account the
  Goldberger-Treiman discrepancy in our analysis. For not too heavy
  pion masses, we expect the uncertainty in the strength of the
  OPEP to be dominated by the current uncertainty in the lattice-QCD
  calculations of $g_A$.}
  $M_\pi$.
Finally, the discontinuity across the  left-hand cuts  is also
affected by the $M_\pi$-dependence of the nucleon mass through its
appearance in the integral equation
(\ref{MeqLOk0integrated}).
To account for the last two  effects in a way  that minimizes the
theoretical uncertainty, we use lattice-QCD results for the
$M_\pi$ dependence of $g_A, F_\pi$ and   $m_N$.   In particular,
we performed quadratic polynomial regression fits (as functions of
$M_\pi^2$) of the lattice-QCD data for pion masses up to $M_\pi =
500\;$MeV  as  shown in Fig.~3 of Ref.~\cite{Baru:2015ira}.
We refer the reader to this paper for more details on the fits and references to
the included lattice-QCD calculations of these quantities.
Using the above results,
we can  generalize the LO LETs by calculating  the  effective range
function  $F(k)$, see Eq.~(\ref{Fk}),
as a function of the (inverse) scattering length at \emph{arbitrary} values
of the pion mass.
Using a single input quantity such as the binding energy or the
scattering length, we can then predict the phase shifts and extract the effective range $r$ and  the
shape parameters $v_i$ (provided the effective
range  function  does  not  have  poles  near  the origin).
Some exemplary  results  for such
predictions  are  discussed in  Ref.~\cite{Baru:2015ira}   for
different  values of the pion mass.
Notice that  contrary  to the chiral extrapolations performed in the
framework of chiral EFT,  no assumptions
about  the  short-range  interaction
$C_0$ as a function of the pion mass 
is  made  when  calculating the LETs.
Instead, we  perform fully independent calculations at each given value of
the pion mass as if we lived in different worlds characterized by a
specific value of $M_\pi$.
Thus, for each considered value of the pion mass,  the LEC $C_0$ has 
to be adjusted to reproduce the given value of the  scattering length used
as input.
By providing relations between various low-energy observables at
unphysical values of the pion mass, the LETs may serve as  
consistency checks of  the  lattice QCD  calculations.

As already emphasized above, the extension of the LETs to NLO is
achieved by including subleading contact interactions 
parametrized via resonance saturation,  see Eq.~(\ref{satur}).  
Retaining the light-quark mass
variation in the subleading contact interaction is formally suppressed
according to the chiral EFT estimates.
On the other hand,  allowing for a variation of this term with $M_\pi$ 
can be used to estimate the theoretical
uncertainty of our analysis. Following Ref.~\cite{Baru:2015ira}, this is achieved by adjusting  the
strength $\beta$ of the short-range interaction  to reproduce the
effective range at the physical point and by assuming
that the $M_\pi^2$-dependence  of $\beta$ is within the envelope
built by the straight lines which go through the physical point and describe a
$\pm 50\%$ change in the value of $\beta$ for $M_\pi = 500$
MeV, i.e.:
\begin{equation}
\label{beta_range}
1 - \delta \beta \left|\frac{M_\pi^2 -(M_\pi^{\rm phys})^2}{\Delta
    M_\pi^2} \right|    \leq \frac{\beta (M_\pi)}{\beta (M_\pi^{\rm
    phys})} \leq 1 + \delta \beta \left|\frac{M_\pi^2 -(M_\pi^{\rm phys})^2}{\Delta
    M_\pi^2} \right|\,,
\end{equation}
with $\delta \beta = 0.5$ and $\Delta M_\pi^2 \equiv (M_\pi^2 -
(M_\pi^{\rm phys})^2 )\big|_{M_\pi = 500\; \mbox{MeV}}$.
Such a choice of $\delta \beta$ is motivated by the fact that it would
cover the known
$M_\pi$-dependence of  $g_A$, $F_\pi$ and $m_N$ if the same
procedure is applied to these quantities. In the next section,
we will also give results corresponding to the more conservative
choice of $\delta \beta = 1.0$.

\vspace*{-0.4cm}
\section{Application of the LETs to the NPLQCD results at $M_\pi\sim 450$ MeV}
\label{sec:LatticeQCD}

As explained in the previous section,  the LETs allow one to perform
consistency checks of lattice-QCD results for NN scattering provided
more than a single observable is extracted.
Unfortunately,  most of the lattice calculations in the NN sector have
so far focused on the determination of the binding energies.
One exception is the work by the NPLQCD collaboration  at the pion
mass of $M_\pi\sim 
800$~MeV 
\cite{Beane:2013br}, which provides, in addition to the binding
energies, also the values of the scattering length, effective range
and even the first shape parameter. It is, furthermore, conjectured in that paper 
that the effective range, expressed in
units of the pion mass, may be approximated by a linear function 
of $M_\pi$.
While  the LETs are certainly beyond their range of applicability
at such heavy pion masses,
this conjecture  was tested  using the LETs  in  our previous work
\cite{Baru:2015ira}, where the resulting $M_\pi $-dependence of the
deuteron binding energy  was indeed found to be in good agreement  with  the
general trend  of lattice data
\cite{Beane:2011iw,Yamazaki:2015asa,Orginos:2015aya,Beane:2013br,Yamazaki:2012hi}.

\begin{figure}[tb]
\vspace*{-0.cm}
\includegraphics[width=0.49\textwidth,keepaspectratio,angle=0,clip]{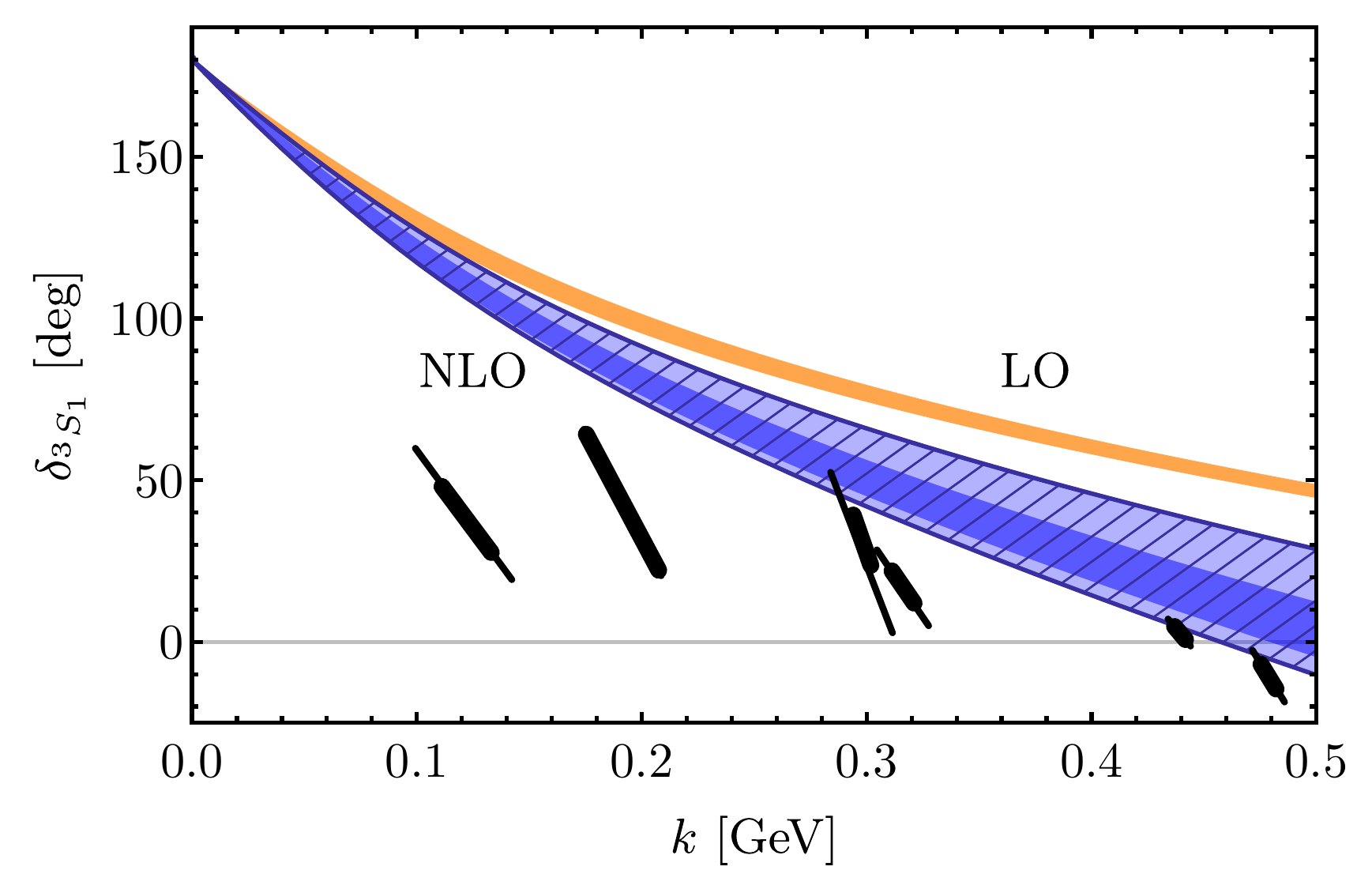}\quad
\includegraphics[width=0.49\textwidth,keepaspectratio,angle=0,clip]{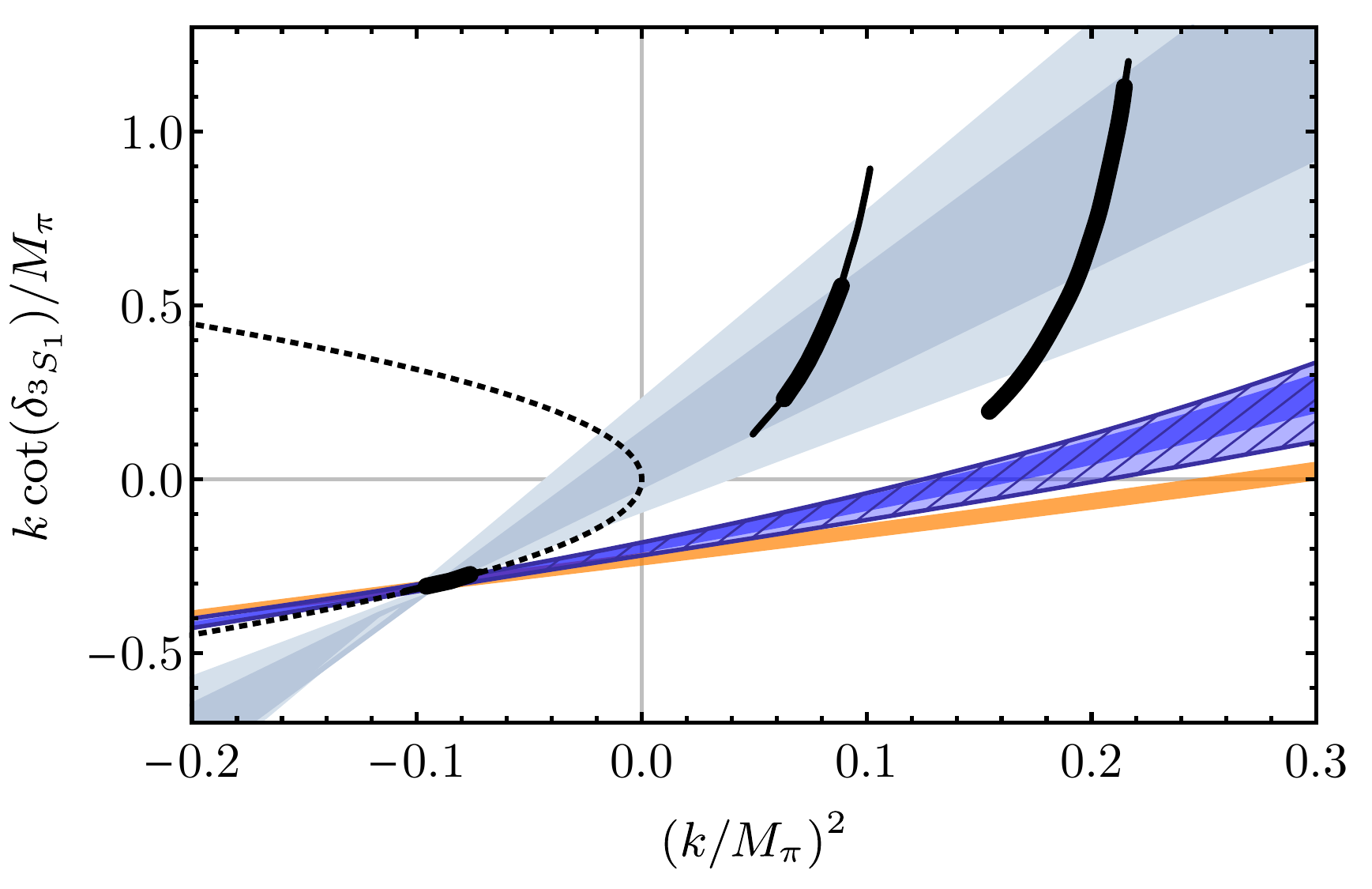}
    \caption{(Color online) Neutron-proton phase shifts (left panel) and the
      effective-range function (right panel) in the  $^3S_1$ channel
      calculated on the lattice at $M_\pi\sim
      450$~MeV~\cite{Orginos:2015aya} 
(filled black regions) in
      comparison with the predictions based on the LETs at LO (orange
      light-shaded bands)
      and NLO (blue dark-shaded and hatched blue light-shaded bands) using the
      NPLQCD result for the deuteron binding energy  $B_d$ as
      input. The uncertainty at LO shown by the orange bands
is entirely given by the uncertainty of $B_d$ in
Eq.~\eqref{BD}. The NLO dark-shaded (hatched light-shaded) bands
correspond to the uncertainty in $B_d$ and the theoretical
uncertainty of the LETs estimated via the variation of $\beta$ with
$\delta \beta = 0.5$ ($\delta \beta = 1.0$) combined in quadrature. 
The grey light- and dark-shaded bands in the right panel depict the
fit results of the lattice points of Ref.~\cite{Orginos:2015aya} based
on the effective range approximation.  The energy of the bound
(virtual) states corresponds to the intersection points of the 
effective-range function $k \cot \delta^{(^3S_1)}$ and the unitarity  term  $ik/M_\pi=\pm
\sqrt{-(k/M_\pi)^2}$, shown by the dotted line in the right panel, in
the lower (upper) half-plane. The phase shift
  corresponds to the Blatt-Biedenharn parametrization of the
S-matrix~\cite{Blatt:1952zz}. 
\label{fig:3S1LET}
}
\end{figure}

Recently,  new results for  NN scattering in the  $^3S_1$  and $^1S_0$
channels have been reported by  the NPLQCD collaboration at $M_\pi\sim
450$~MeV~\cite{Orginos:2015aya}.  The calculations were performed  for
$n_f=2+1$  flavors  of light quarks  at three  lattice volumes  of
$L=2.8$~fm,  $L=3.7$~fm,  and  $L=5.6$~fm  using the lattice spacing
of $b=0.12$~fm.
In analogy to their previous work, the scattering phase shifts for
the  $^3S_1$  and $^1S_0$ partial waves were extracted  for several
values of the cms NN momenta using the extended L\"uscher
approach~\cite{Luscher:1986pf,Luscher:1990ux,Briceno:2013bda}
as shown by the black filled regions in  Fig.~\ref{fig:3S1LET} for the
case of the $^3S_1$ channel.
In addition  to the phase shifts,  the binding energies
of the deuteron and the  dineutron were extracted.
Thus, it is interesting to test whether these results fulfill the LETs
introduced above.

\subsection{The $^3S_1$ channel}
\label{sec:3s1}

The deuteron binding energy calculated in
Ref.~\cite{Orginos:2015aya}  at $M_\pi\simeq 450$ MeV at  three
lattice volumes  and  extrapolated to the infinite volume is
\begin{equation}\label{BD}
B_d=14.4 \big({}^{+3.2}_{-2.6} \big)\  {\rm MeV },
\end{equation}
where the errors include
statistical and systematic uncertainties as well as the
extrapolation  uncertainty  combined in quadrature.
Further, the first two coefficients in the effective range expansion, namely the scattering length
and the effective range,  were determined in
Ref.~\cite{Orginos:2015aya} by fitting the effective range approximation
of the effective range function,
\begin{equation}
\label{eq:ere}
k \cot \delta \simeq - \frac{1}{a} + \frac{1}{2} r k^2\,,
\end{equation}
to the two lowest-energy scattering data points and the deuteron
binding  energy, see the grey bands in the right panel of Fig.~\ref{fig:3S1LET}.
Notice that all three lattice data correspond to
nucleon momenta below the branch point $| k| = M_\pi/2$  of the left-hand
cut from the OPEP.  The resulting values for the inverse scattering
length and the effective range in units of the pion mass reported in
Ref.~\cite{Orginos:2015aya} are
\begin{eqnarray}
\label{ERE3s1}
\big( M_\pi a^{(^3 \hskip -0.025in S _1)} \big)^{-1}  =  
  -0.04\big({}^{+0.07}_{-0.10}\big) \big({}^{+0.08}_{-0.17} \big),  \quad \quad  M_\pi r^{(^3 \hskip
  -0.025in S _1)} \ =\
7.8 \big({}^{+2.2}_{-1.5} \big) \big({}^{+3.5}_{-1.7} \big)\,,
\end{eqnarray}
where the uncertainties in the first and second brackets are  statistical  and systematic, respectively.

In Fig.~\ref{fig:3S1LET},  we confront the lattice-QCD phase shifts of
Ref.~\cite{Orginos:2015aya} with
the predictions of the LETs  at LO  and NLO.
We use the NPLQCD result for the deuteron binding energy given in
Eq.~\eqref{BD} as input to adjust  the
leading-order contact term $C_0$.   This is sufficient to predict the
phase shift at LO. As explained in the previous section, there are no
additional parameters at NLO. As shown in the left panel 
of Fig.~\ref{fig:3S1LET}, the change in the phase shifts  when going  from
 LO to NLO is reasonably small which confirms a good
convergence of the LETs in this channel. The
expected accuracy of the NLO prediction can be roughly estimated by
the width of the blue band generated by the variation of the parameter
$\beta$ as described above and appears to be consistent with the shift from LO to
NLO. Notice that the LO (orange) band reflects the uncertainty in the
NPLQCD prediction of the binding energy and does not
include the theoretical uncertainty of the LETs.  As  required by
the Levinson theorem for  the case of a bound deuteron, the  phase shifts  generated by
the LETs  go through  $180^\circ$  at the  origin.
Comparing the NPLQCD results for phase shifts and the effective range
function in the $^3S_1$ channel with those based on the LETs as
visualized in Fig.~\ref{fig:3S1LET}, we end up with the following
conclusions:
\begin{itemize}
\item
First, as shown in the right panel of Fig.~\ref{fig:3S1LET}, only
\emph{positive} values of the
scattering length appear to be consistent with the NPLQCD result for the
deuteron binding energy quoted in Eq.~(\ref{BD})
as opposed to the negative central value for $a^{(^3 \hskip -0.025in S
  _1)}$ reported in Ref.~\cite{Orginos:2015aya}. Our results for the inverse scattering length
extracted from $B_d$ by means of the LETs disagree with the NPLQCD
ones given in Eq.~(\ref{ERE3s1})  as can be
inferred  from the right panel of Fig.~\ref{fig:3S1LET}.
\item
While the lattice phase shifts at higher momenta are in  reasonable
agreement with the ones predicted by the LETs, their low-momentum
behavior is incompatible (within the quoted errors) with that
predicted by the LETs as demonstrated in  both panels in  Fig.~\ref{fig:3S1LET}.
In particular, the phase shift  calculated on the lattice at the
lowest considered momentum of $k \simeq 122$ MeV,  $\delta =
38\big({}^{+13}_{-11}\big) \big({}^{+23}_{-16}\big)$
degrees,     is   a  factor of  three
smaller than  the corresponding value of $\delta = 111 (\pm
5)$ degrees extracted  from the LETs. 
\item
An extrapolation of the lattice data to  zero momenta
in the left panel of  Fig.~\ref{fig:3S1LET} seems to indicate 
that the phase shift goes to zero.  This would, however, contradict
the existence of a bound state in this partial wave as  a
consequence of  the Levinson theorem (or require shifting
$\delta_{3S1} $ by $180$~degrees in the entire plotted energy range
which would be inconsistent with the LETs).
\end{itemize}
One may raise a question whether the observed inconsistencies between the
lattice-QCD results for phase shifts and the LETs predictions
could originate from underestimating the quark mass dependence of the
NLO contact interaction by constraining the function $\beta (M_\pi )$ as
described in the previous section.  To clarify this issue, we have
increased the allowed variation of $\beta$ by a factor of two, i.e.~we set $\delta \beta = 1$
instead of $\delta \beta = 0.5$. This corresponds to the allowed
variation of the strength of the short-range term at $M_\pi = 500$~MeV
by $\pm 100\%$ as compared to its value at the physical
point. 
The resulting predictions for the phase shifts and the effective range
function are shown by
the hatched blue light-shaded bands 
in  Fig.~\ref{fig:3S1LET}. 
With the resulting uncertainty nearly covering the shift from our LO to
NLO results, we expect such an error estimation to be too
conservative. Still, none of our conclusions appear to be affected by employing
this very conservative uncertainty estimation.

\begin{figure}[tb]
\vspace*{-0.cm}
\includegraphics[width=0.7\textwidth,keepaspectratio,angle=0,clip]{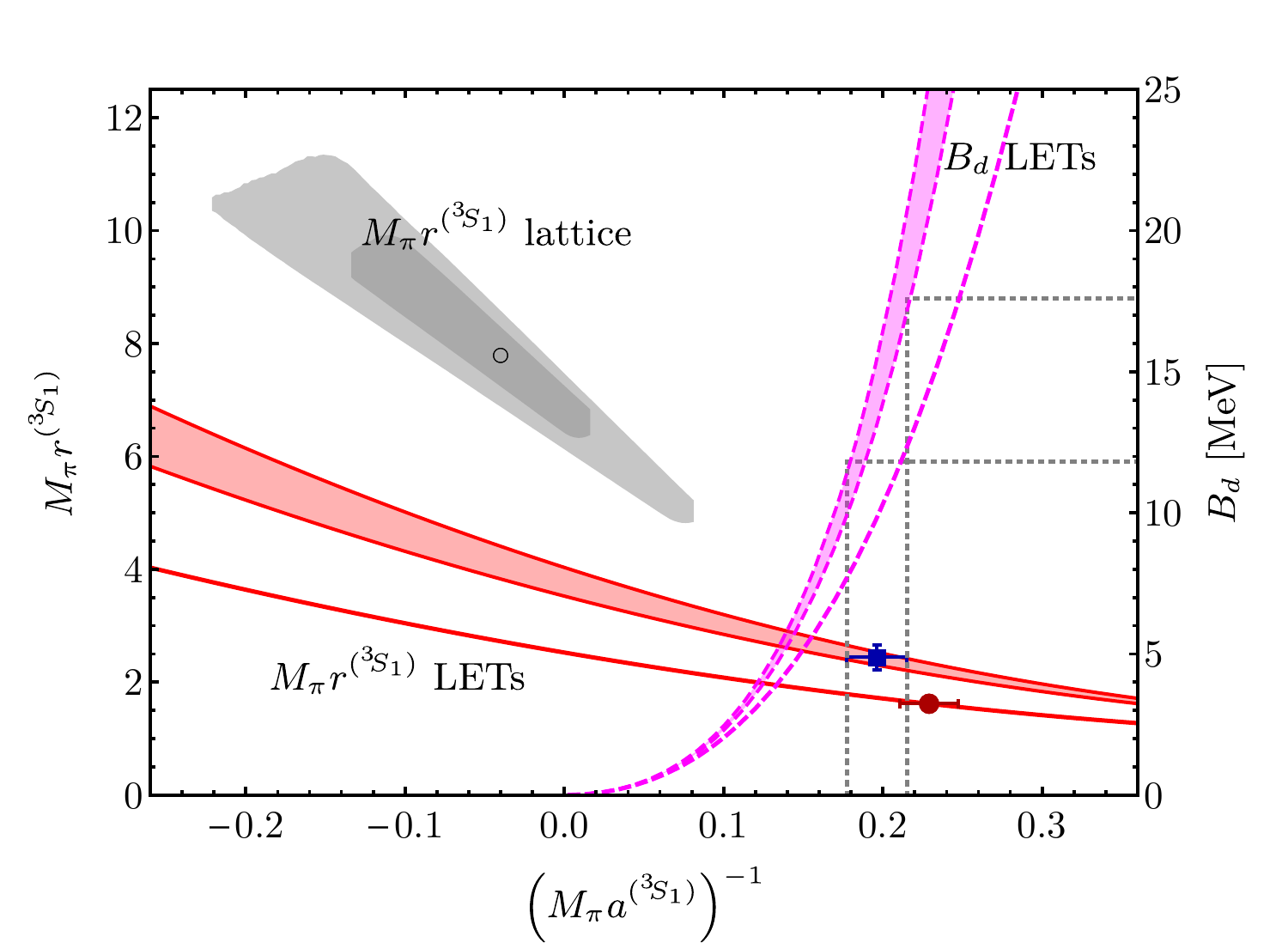}
    \caption{(Color online) Correlations between the inverse scattering length
      $a^{-1}$, effective range $r$ and the binding energy in the
      $^3S_1$ partial wave induced by the one-pion
exchange potential. The red solid and dashed magenta lines
show the predictions of the LO LETs for $M_\pi
r^{(^3S_1)}$ and $B_d$. The light-shaded bands between the red solid
and dashed magenta lines visualize the predictions of the NLO LETs for $M_\pi
r^{(^3S_1)}$ and $B_d$, respectively, and reflect the theoretical
uncertainty estimated via the variation of $\beta$ with
$\delta \beta = 0.5$ as described in the text. 
The horizontal dotted lines
specify the range of values for $B_d$
consistent  with the lattice-QCD results of
Ref.~\cite{Orginos:2015aya} for this observable.
The solid dark-red circle (blue rectangle) shows the LO (NLO) LET
predictions for the effective range.
The open black circle gives the result for the inverse scattering length and
effective range reported by the
  NPLQCD collaboration~\cite{Orginos:2015aya} while the grey area
  around it shows the estimated uncertainty from that paper. All results
  correspond to the Blatt-Biedenharn parametrization of the
S-matrix~\cite{Blatt:1952zz}.
\label{fig:LETs_3s1}
}
\end{figure}

We are now in the position to employ the LETs in order to
extract  the  scattering length and the
effective range from the deuteron binding energy calculated by the
NPLQCD collaboration.  Such an extraction is  possible
because the effective range
function does not  possess poles  at low momenta  and,  therefore, can
be Taylor expanded around the origin\footnote{The  effective range
  function does have a pole at  $k\simeq 500$ MeV  where the phase
  shift crosses zero but  these  momenta are already beyond the region
  of  the validity  of the effective range expansion.}.
In Fig.~\ref{fig:LETs_3s1},  we plot the  deuteron binding energy and
the  effective range as functions of the inverse scattering length in
units of the pion mass predicted by the LETs at LO (shown by the
lines) and NLO (shown by the bands).
Specifically, the red band between two solid lines  represents  the
NLO LET calculation for  the effective range  as a function  of the
inverse scattering length.  Similarly,  the magenta  band  between two
dashed lines shows the  deuteron binding energy  versus the inverse
scattering length at NLO.  Further,  the two horizontal  dotted lines
separate the region  of  the binding energies
consistent with  the NPLQCD result of Ref.~\cite{Orginos:2015aya},  Eq.~\eqref{BD},
 for  the binding energy.  Projecting  this area onto the $x$-axis,  as
 shown by  the  vertical lines,  one obtains the corresponding values of  the
 scattering length and the effective range from the LETs.
In particular, we find
\begin{eqnarray}\label{ERENLO}
\big( M_\pi a_{\rm LET, \, LO}^{(^3 \hskip -0.025in S _1)} \big)^{-1}  &=&
0.229 \big({}_{-0.018}^{+0.019} \big), \quad \quad \mbox{\hspace{1.4cm}}
\! M_\pi r_{\rm LET, \, LO}^{(^3 \hskip -0.025in S _1)} \ =\
1.62 \big({}_{-0.06}^{+0.06} \big)
\,, \nn
\big( M_\pi a_{\rm LET, \, NLO}^{(^3 \hskip -0.025in S _1)} \big)^{-1}  &=&
0.196 \big({}^{+0.014}_{-0.013} \big) \big({}^{+0.007}_{-0.004} \big) 
, \quad \quad
\! M_\pi r_{\rm LET, \, NLO}^{(^3 \hskip -0.025in S _1)} \ =\
2.44 \big({}^{+0.08}_{-0.08} \big) \big({}^{+0.12}_{-0.17} \big) \,,
\end{eqnarray}
which correspond to the following values in units of fm
\begin{eqnarray}\label{ERENLOfm}
a_{\rm LET, \, LO}^{(^3 \hskip -0.025in S _1)}  &=&
1.915\big({}^{+0.159}_{-0.147} \big)  \text{ fm}, \quad \quad \mbox{\hspace{1.4cm}}
\! r_{\rm LET, \, LO}^{(^3 \hskip -0.025in S _1)} \ =\
0.71\big({}^{+0.02}_{-0.03} \big)   \text{ fm}\,, \nn
a_{\rm LET, \, NLO}^{(^3 \hskip -0.025in S _1)}   &=&
2.234\big({}^{+0.156}_{-0.144} \big) \big({}^{+0.052}_{-0.072} \big) 
\text{ fm},
\quad \quad
\! r_{\rm LET, \, NLO}^{(^3 \hskip -0.025in S _1)} \ =\
1.07 \big({}^{+0.03}_{-0.03} \big) \big({}^{+0.05}_{-0.08} \big) \text{ fm}.
\end{eqnarray}
Here, the errors in the first brackets reflect the uncertainty in
the value of the deuteron binding energy in Eq.~(\ref{BD}) used as
input. For the NLO results, we also give in the second brackets an
estimation of the theoretical uncertainty corresponding to the choice
of $\delta \beta = 0.5$.  
Clearly, the above values are at variance with those extracted by the
NPLQCD collaboration  and given in
Eq.~(\ref{ERE3s1}).  In
particular, our value for  the effective range is about a factor
of $3$ smaller than the one found
in Ref.~\cite{Orginos:2015aya}.  Interestingly, the NLO LET prediction
for the effective range is in
excellent  agreement with the assumed linear in $M_{\pi}$ behavior
of the quantity  $M_\pi r^{(^3 \hskip -0.025in S _1)}$ conjectured  in Ref.~\cite{Beane:2013br},
cf. Fig.~\ref{fig:3s1eff_rad_mpi} and the right panel of Fig.~\ref{fig:eff_rad_mpi}.
For the sake of completeness, we also give the NLO LET results based
on a more conservative uncertainty estimation resulting by employing
a weaker constraint on the allowed $M_\pi$-dependence of the subleading contact
interaction corresponding to the choice of $\delta
\beta =1$: 
\begin{equation}\label{ERENLOCons}
\big( M_\pi a_{\rm LET, \, NLO}^{(^3 \hskip -0.025in S _1)} \big)^{-1}  =
0.196 \big({}^{+0.014}_{-0.013} \big) \big({}^{+0.018}_{-0.008} \big) 
, \quad \quad
\! M_\pi r_{\rm LET, \, NLO}^{(^3 \hskip -0.025in S _1)} \ =\
2.44 \big({}^{+0.08}_{-0.08} \big) \big({}^{+0.21}_{-0.47} \big) \,,
\end{equation}
or
\begin{equation}\label{ERENLOfmCons}
a_{\rm LET, \, NLO}^{(^3 \hskip -0.025in S _1)}   =
2.234\big({}^{+0.156}_{-0.144} \big) \big({}^{+0.093}_{-0.191} \big) 
\text{ fm},
\quad \quad
\! r_{\rm LET, \, NLO}^{(^3 \hskip -0.025in S _1)} \ =\
1.07 \big({}^{+0.03}_{-0.03} \big) \big({}^{+0.09}_{-0.21} \big) \text{ fm}.
\end{equation}
in units of fm. 

To understand the origin of the disagreement between our results for
the scattering length and effective range with those of
Ref.~\cite{Orginos:2015aya}, 
it is instructive to take a closer look at the procedure for
their determination employed by the NPLQCD collaboration. 
To this aim, a fit of the lattice phase-shift data at the two lowest
energies and the deuteron pole
was performed using the 
effective-range
approximation~(\ref{eq:ere}).
Note that the considered phase-shifts and the deuteron pole correspond  to momenta    below the
branch point  of the $t$-channel cut due to the OPEP. 
For non-singular
potentials of a finite range, the applicability region of the effective range
expansion is given by the inverse range of the interaction which
determines the position of the first left-hand singularity.
Consequently, the effective range and shape parameters may be expected
to scale with the corresponding powers of the pion mass. For example, 
for the physical value of the pion mass, one has $r^{(^1S_0)} = 1.9 \,
M_\pi^{-1}$ and  $r^{(^3S_1)} = 1.2 \, M_\pi^{-1}$. The very large
value of the effective range 
reported by the NPLQCD collaboration, $r^{(^3S_1)} = 7.8 \,
M_\pi^{-1}$, 
either indicates that the range of the nuclear force is 
considerably larger than that of
the OPEP or signals the appearance of a pole in the effective-range
function in the near-threshold region.\footnote{For example, such a
  pole very close to threshold appears  in the spin-doublet $S$-wave channel for neutron-deuteron
scattering.} In both cases, the applicability range of the 
effective range expansion of $k \cot \delta$ would be significantly
smaller than one may expect based on the position of the 
left-hand cut due to the OPEP. As a consequence, the
solution for $a^{(^3S_1)}$ and  $r^{(^3S_1)}$ reported in Ref.~\cite{Orginos:2015aya}
and listed in Eq.~(\ref{ERE3s1}) is not self-consistent in the sense
that it is obtained by fitting the effective range approximation to the data points
outside of its validity region which can be roughly estimated as $| k |
\lesssim 2/ r^{(^3S_1)} \sim 0.26 \, M_\pi$. Specifically, the
deuteron binding momentum at $M_\pi \simeq 450$~MeV is of the order of
$\gamma \sim 0.3 \, M_\pi$, whereas the  phase-shifts data employed in the  analysis correspond to  $k\sim 0.27\, M_\pi$  and  $k\sim 0.42\, M_\pi$.

To get further insights into this
issue, consider the two roots of the quadratic equation
$-1/a + r k^2/2 -ik =0$
which determines
the pole positions of the scattering amplitude within the
effective-range approximation,
\begin{equation}
k_1 = \frac{i}{r}  \bigg( 1 + \sqrt{1-\frac{2r}{a}}  \bigg) \simeq
i  \bigg(\frac{2}{r} - \frac{1}{a} \bigg) \,, \quad \quad
k_2 = \frac{i}{r}  \bigg( 1 - \sqrt{1-\frac{2r}{a}}  \bigg) \simeq
\frac{ i }{a} \bigg(1+ \frac{r}{2a}       \bigg)
\,,
\end{equation}
where we have expanded the square root in powers of $r/a$ and
neglected terms of order $\mathcal{O} \big( (r/a )^2 \big)$. This is
justified both for the physical value of the pion mass and
for the solution given in Eq.~(\ref{ERE3s1}), since in both cases one has  $|r/a|
\sim 0.3$.  At the physical pion mass, the second root yields the
deuteron binding momentum $k_2\simeq 45i$ MeV while the first root,
$k_1\simeq 200i$ MeV,  lies outside of the applicability region of the
 effective range expansion and is an artifact of the effective range
 approximation.  In particular, it disappears or changes the position
 upon including higher-order terms in the effective range
 expansion. On the contrary, for the solution in Eq.~(\ref{ERE3s1}) at
 $M_{\pi}\simeq 450$~MeV, the deuteron
 pole corresponds to the first root,
 $k_1\simeq 135i$~MeV,  where  the  dominant contribution comes
 from the effective range.  Meanwhile, because the scattering length
 in  Eq.~\eqref{ERE3s1}  is negative,  the second root corresponds to
 the momentum, $k_2\simeq - 15i$ MeV,  lying on the imaginary axis in the lower half  plane.
 Therefore, the  results  of  Ref.~\cite{Orginos:2015aya} imply the
 existence of a shallow  virtual  state with the excitation energy
 less than $0.5$~MeV in addition to the deuteron, which is not 
supported by our analysis based on the LETs.

\begin{figure}[tb]
\vspace*{-0.cm}
\includegraphics[width=0.49\textwidth,keepaspectratio,angle=0,clip]{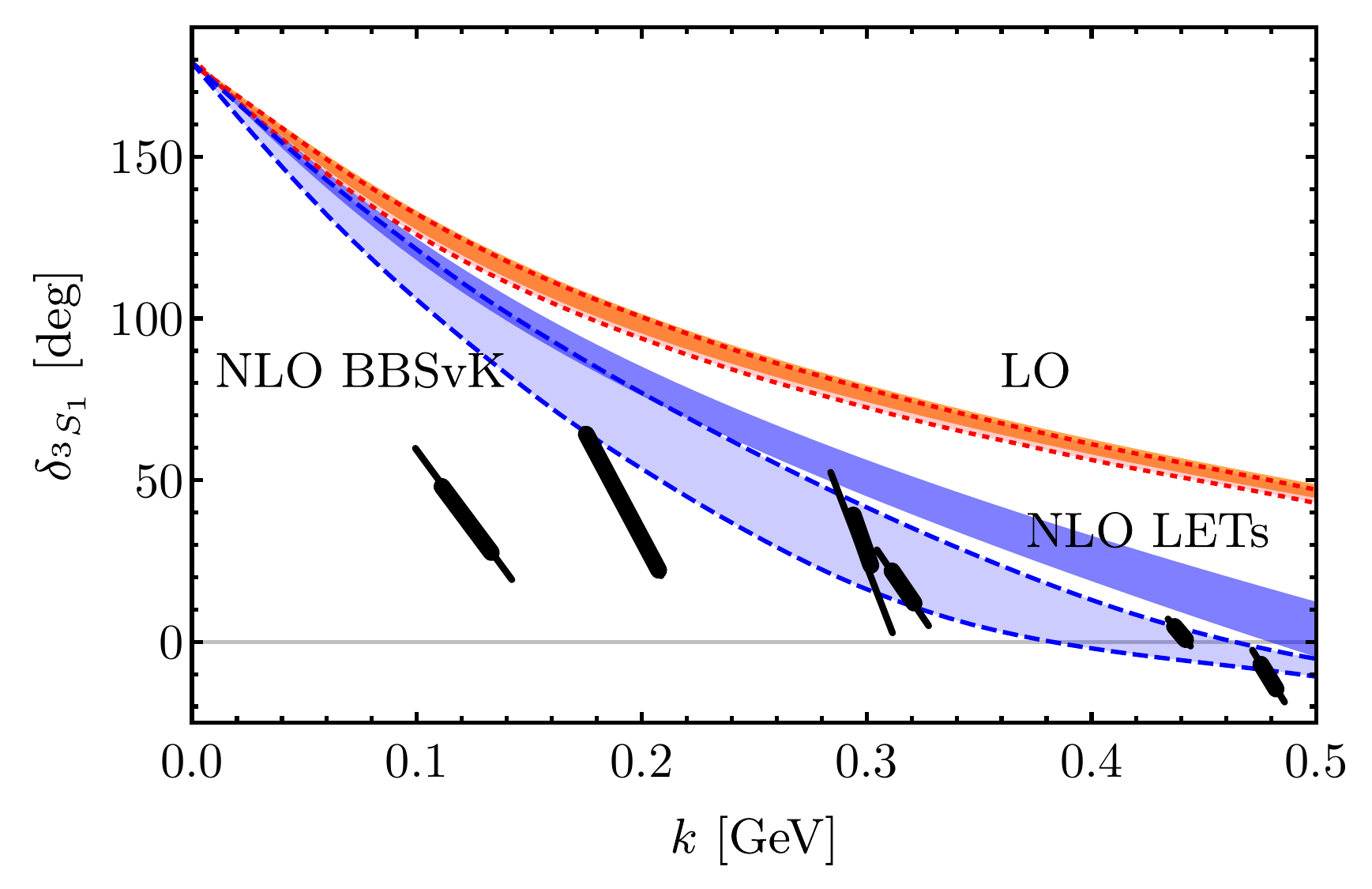}
\includegraphics[width=0.49\textwidth,keepaspectratio,angle=0,clip]{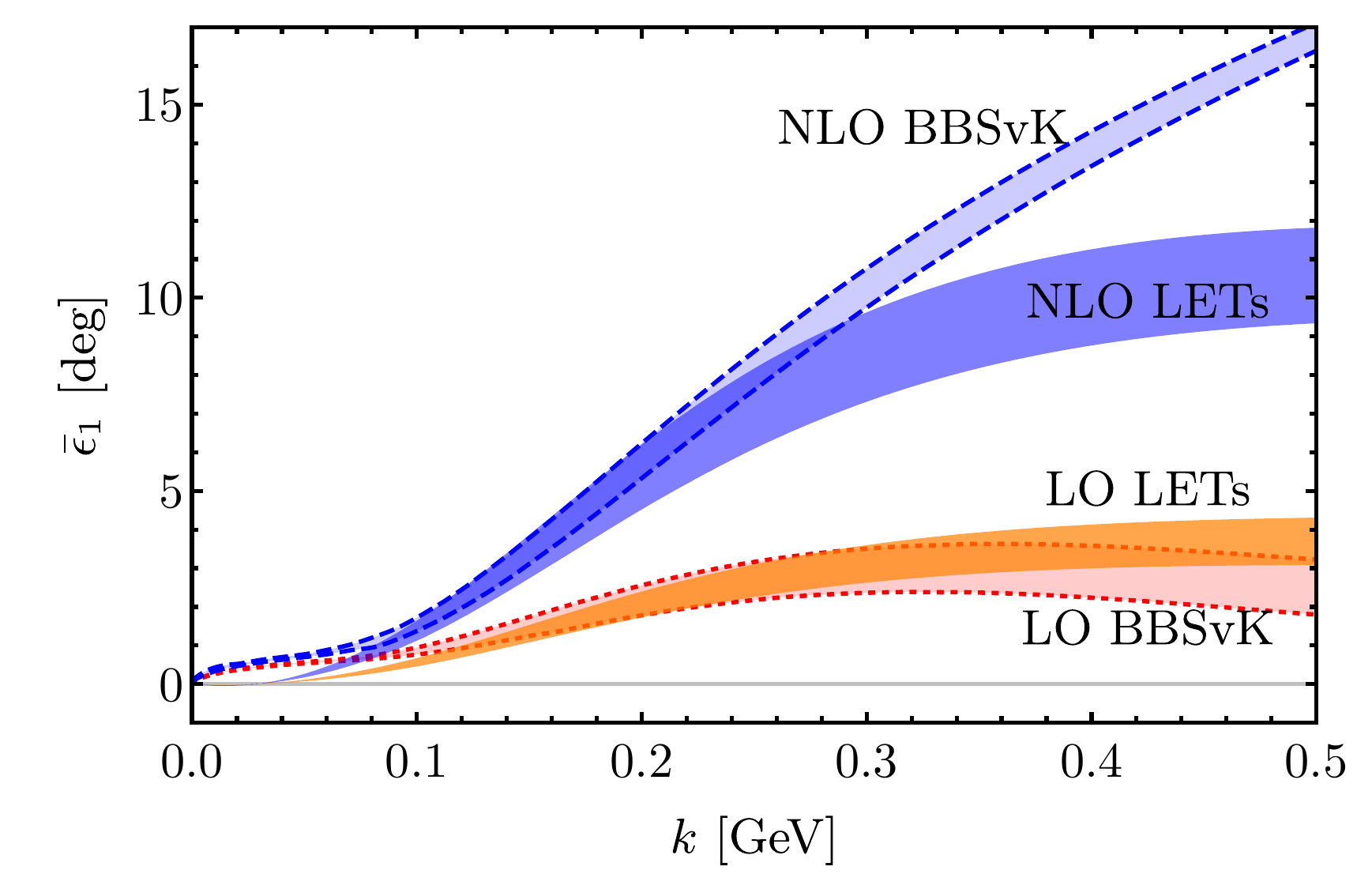}
    \caption{
\label{fig:comp_3s1} (Color online) Neutron-proton phase shifts in the
$^3S_1$ channel (left panel) and the mixing angle $\bar \epsilon_1$ (right
panel)  at $M_\pi\sim 450$~MeV based on the LETs at LO and NLO in
comparison with the results obtained in Ref.~\cite{Orginos:2015aya}
using the EFT formulation of Ref.~\cite{Beane:2001bc}, labelled as BBSvK, 
at LO (light-shaded band between pink dotted lines) and NLO 
(light-shaded band between   blue dashed lines). 
The results for the $^3$S$_1$ phase shift 
  correspond to the Blatt-Biedenharn parametrization of the
S-matrix~\cite{Blatt:1952zz}  while the mixing parameter is shown for
the Stapp parameterization to allow for the comparison with the
results of Ref.~\cite{Orginos:2015aya}. 
For remaining notation see Fig.~\ref{fig:3S1LET}. 
 } 
\end{figure}

Finally, it is interesting to compare our results based on
the LETs
with the ones obtained using an alternative  approach proposed in
Ref.~\cite{Beane:2001bc}, which will be referred to as BBSvK,
where  the expansion of the nuclear force around the chiral limit was
employed, see, however, Ref.~\cite{Epelbaum:2009sd} for a criticism.
This approach was used in Ref.~\cite{Orginos:2015aya}  to
calculate the phase shifts and the mixing angle in  the $^3S_1-{}^3D_1$
channel. A comparison of results from the two approaches is presented
in Fig.~\ref{fig:comp_3s1}.
While the $^3S_1$ phase
shift and the mixing angle show a very similar behavior at LO, there
are more sizable differences at NLO.
Notice that apart from the different treatment of pions, the two
approaches also differ in the way the NLO
short-range interaction is taken into account.  In particular,
in Ref.~\cite{Orginos:2015aya},  the strength of this subleading
short-range term was adjusted to fit the  lattice phase shifts. In
contrast, in our approach, the  strength of the subleading contact
interaction $\beta$ is determined by the value of the
effective range at the physical point while its allowed $M_\pi$  dependence at
unphysical pion masses is used to estimate the theoretical
uncertainty as explained in Sec.~\ref{general}.  This procedure
ensures that both the LO and NLO LET results depend on a single
unknown parameter.  We further emphasize that the low-energy behavior
of the mixing angle found in Ref.~\cite{Orginos:2015aya} and shown in
the right panel of Fig.~\ref{fig:comp_3s1} seems to be at variance
with the expected threshold behavior for this quantity, $\bar \epsilon_1
\sim k^3$ (for details see, e.g. \cite{deSwart:1995ui}).
Irregardless of these differences, the two approaches yield
similar numerical results for the $^3S_1$ phase shift and the mixing
angle $\bar \epsilon_1$ in the considered range of momenta.  The
values of the
scattering length and effective range extracted in
Ref.~\cite{Orginos:2015aya} from the lattice data using the framework
of Ref.~\cite{Beane:2001bc} read
\begin{eqnarray}
a_{\rm BBSvK, \, LO}^{(^3S_1)} &=& 1.94(09)(17)\mbox{ fm}, \quad \quad
                                   r_{\rm BBSvK, \, LO}^{(^3S_1)} =
                                   0.674(17)(29)\mbox{ fm}, \nn
a_{\rm BBSvK, \, NLO}^{(^3S_1)} &=& 2.72(22)(27)\mbox{ fm}, \quad \quad
                                   r_{\rm BBSvK, \, NLO}^{(^3S_1)} =
                                   1.43(12)(13)\mbox{ fm},
\end{eqnarray}
where the uncertainties in the first and second brackets correspond to
the statistical  and systematic uncertainties of the lattice results.
As already pointed out, the LO values are in agreement with our LO
predictions given in Eq.~(\ref{ERENLOfm}), while the deviations at NLO and, in
particular, the large value of the effective range are presumably
caused by an attempt to reproduce the lattice-QCD result for the
$^3S_1$ phase shift at $k \simeq 0.2$~GeV within the BBSvK approach.

\subsection{The $^1S_0$ channel}
\label{sec:1s0}

\begin{figure}[tb]
\vspace*{-0.cm}
\includegraphics[width=0.49\textwidth,keepaspectratio,angle=0,clip]{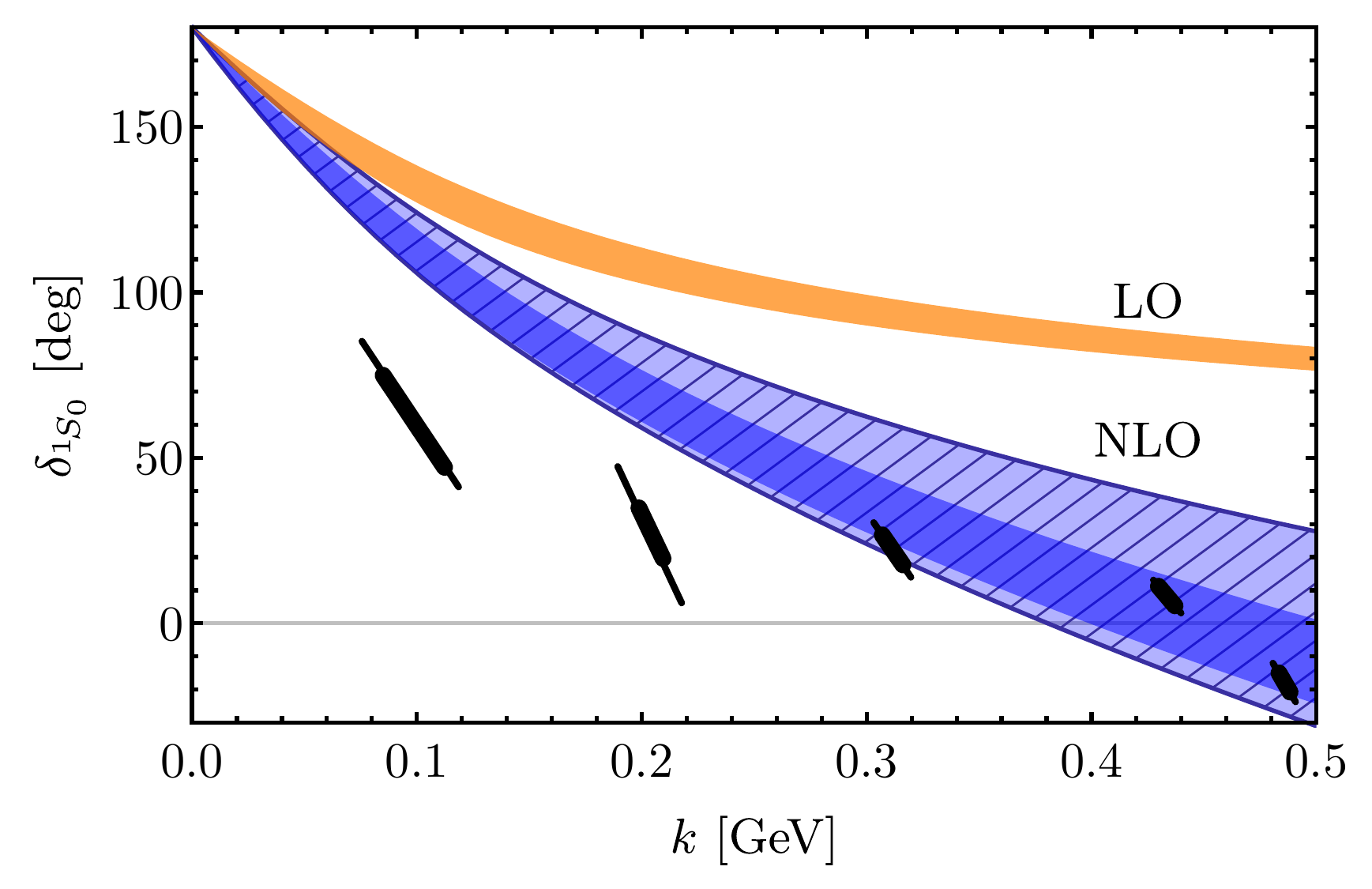}
\includegraphics[width=0.49\textwidth,keepaspectratio,angle=0,clip]{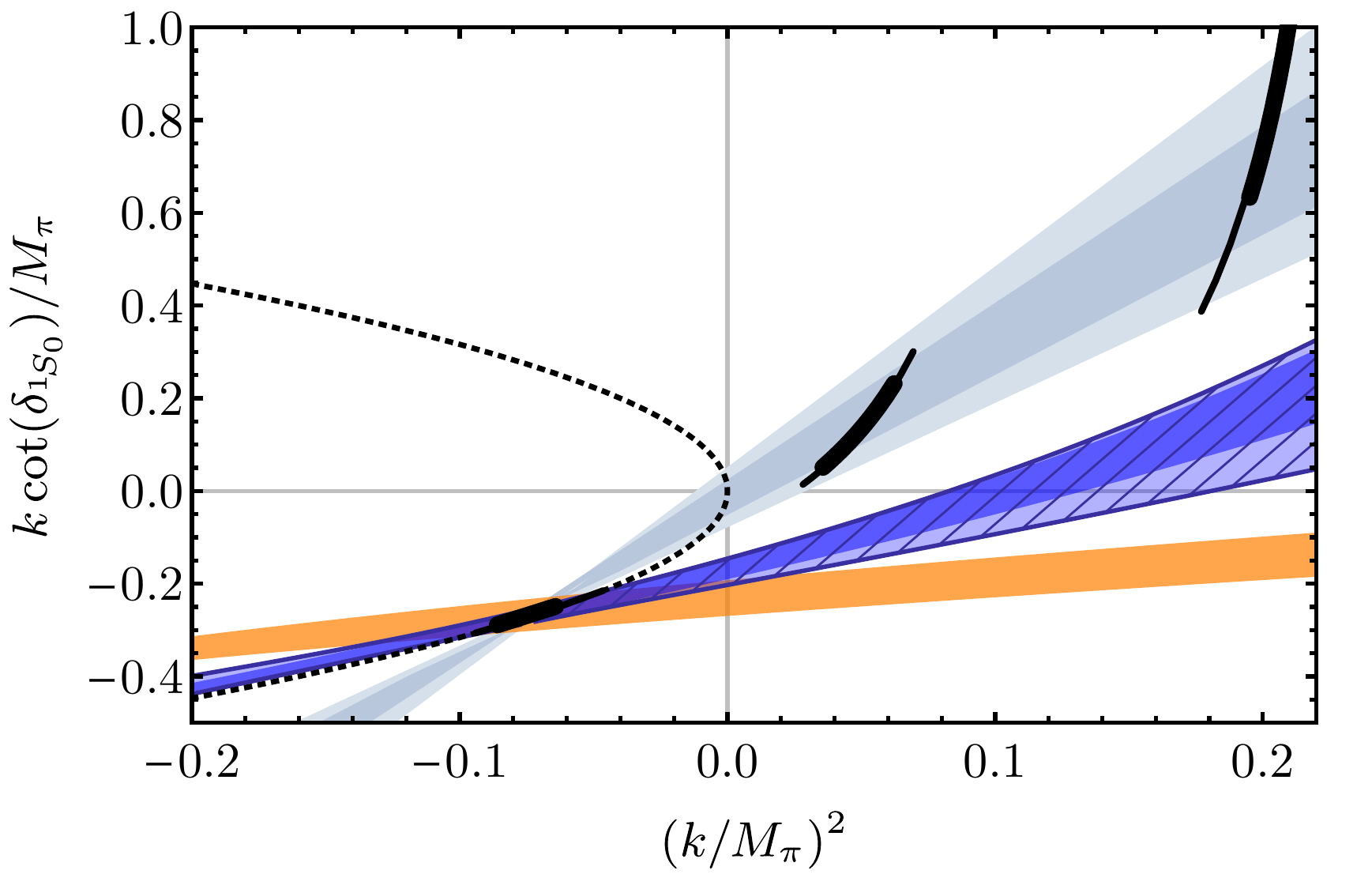}
    \caption{(Color online) Two-nucleon phase shifts (left panel) and the
      effective-range function (right panel) in the  $^1S_0$ channel
      calculated on the lattice at $M_\pi\sim
      450$~MeV~\cite{Orginos:2015aya} in
      comparison with the predictions based on the LETs at LO and NLO using the
      NPLQCD result for the dineutron binding energy  $B_{nn}$ as
      input. For notation see Fig.~\ref{fig:3S1LET}. 
\label{fig:1s0_LETsEB}
}
\end{figure}

We now turn to the spin-singlet channel.
In Fig.~\ref{fig:1s0_LETsEB}, 
we confront the phase shifts extracted based on the LETs 
with the  lattice-QCD results for the $^1S_0$ partial wave. Here we
apply the same procedure as in the $^3S_1$ channel and use
the NPLQCD result for the dineutron binding energy,
\cite{Orginos:2015aya} 
\begin{equation}
\label{EnnNPLQCD}
B_{nn} = 12.5 \big({}^{+3.0}_{-5.0} \big) \mbox{ MeV}\,,
\end{equation}
as input to fix the short-range interaction at LO. The NLO
short-range interaction is again taken into account by means of
resonance saturation, see Eq.~(\ref{satur}), with the strength $\beta$
being determined by the effective range at the
physical point. The allowed $M_\pi$-dependence of $\beta$ is specified
by Eq.~(\ref{beta_range}), and the blue dark-shaded bands in
Fig.~\ref{fig:1s0_LETsEB} correspond to the choice $\delta \beta
=0.5$.  Notice that the shift in the predictions when going from LO to
NLO is now much larger than in the spin-triplet channel which
is in line with the lower predictive power of the LETs in the $^1S_0$
partial wave. Consequently, we believe that a variation of the
strength $\beta$ with $\delta \beta = 0.5$ does  not provide a
realistic estimation of the theoretical uncertainty at NLO in this
channel. To have a more conservative estimation, we will allow for a
larger $M_\pi$-dependence in this channel and set $\delta \beta = 1$
as visualized by the hatched blue light-shaded bands in Fig.~\ref{fig:1s0_LETsEB}.    

\begin{figure}[tb]
\vspace*{-0.cm}
\includegraphics[width=0.49\textwidth,keepaspectratio,angle=0,clip]{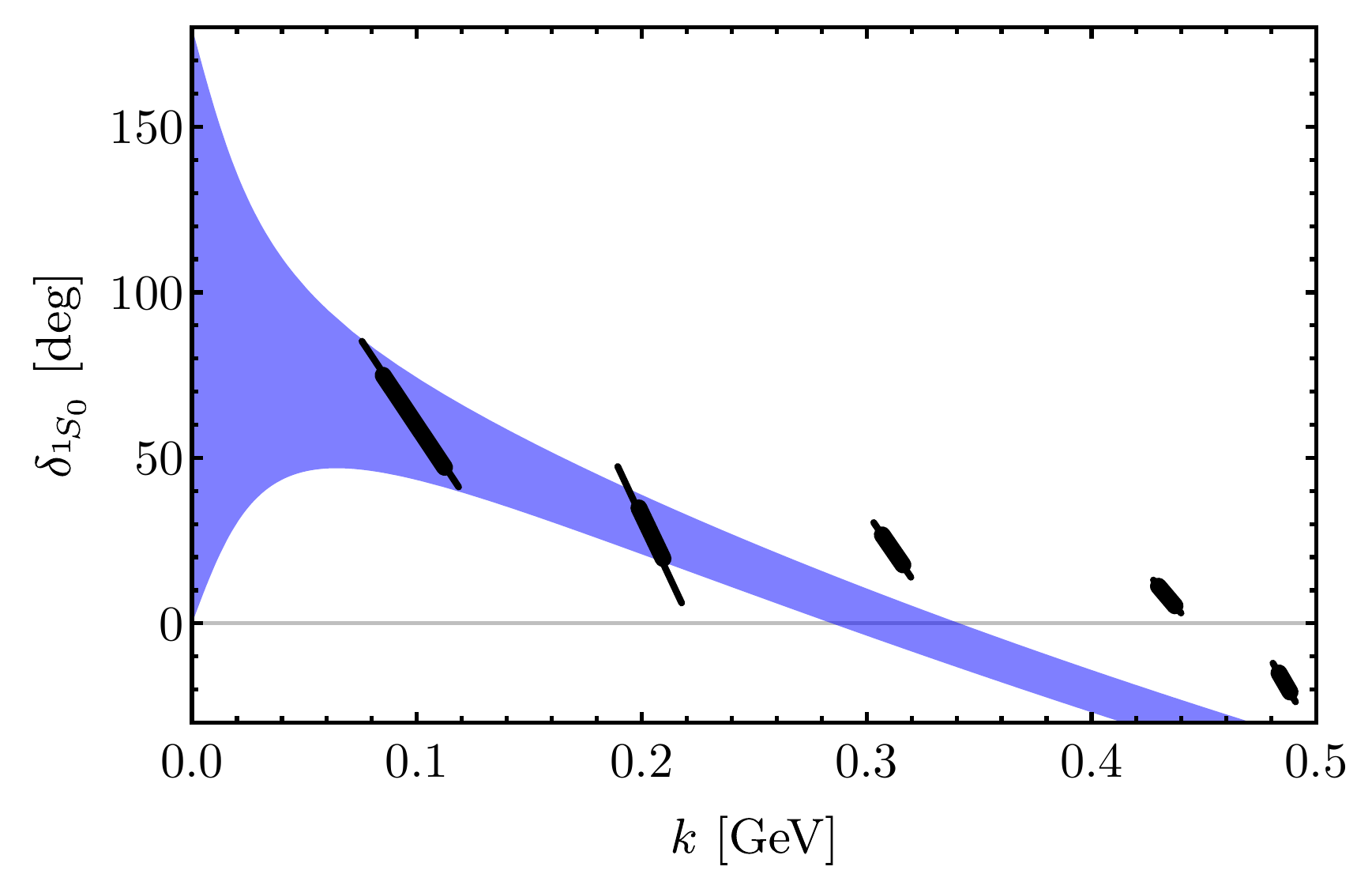}
\includegraphics[width=0.49\textwidth,keepaspectratio,angle=0,clip]{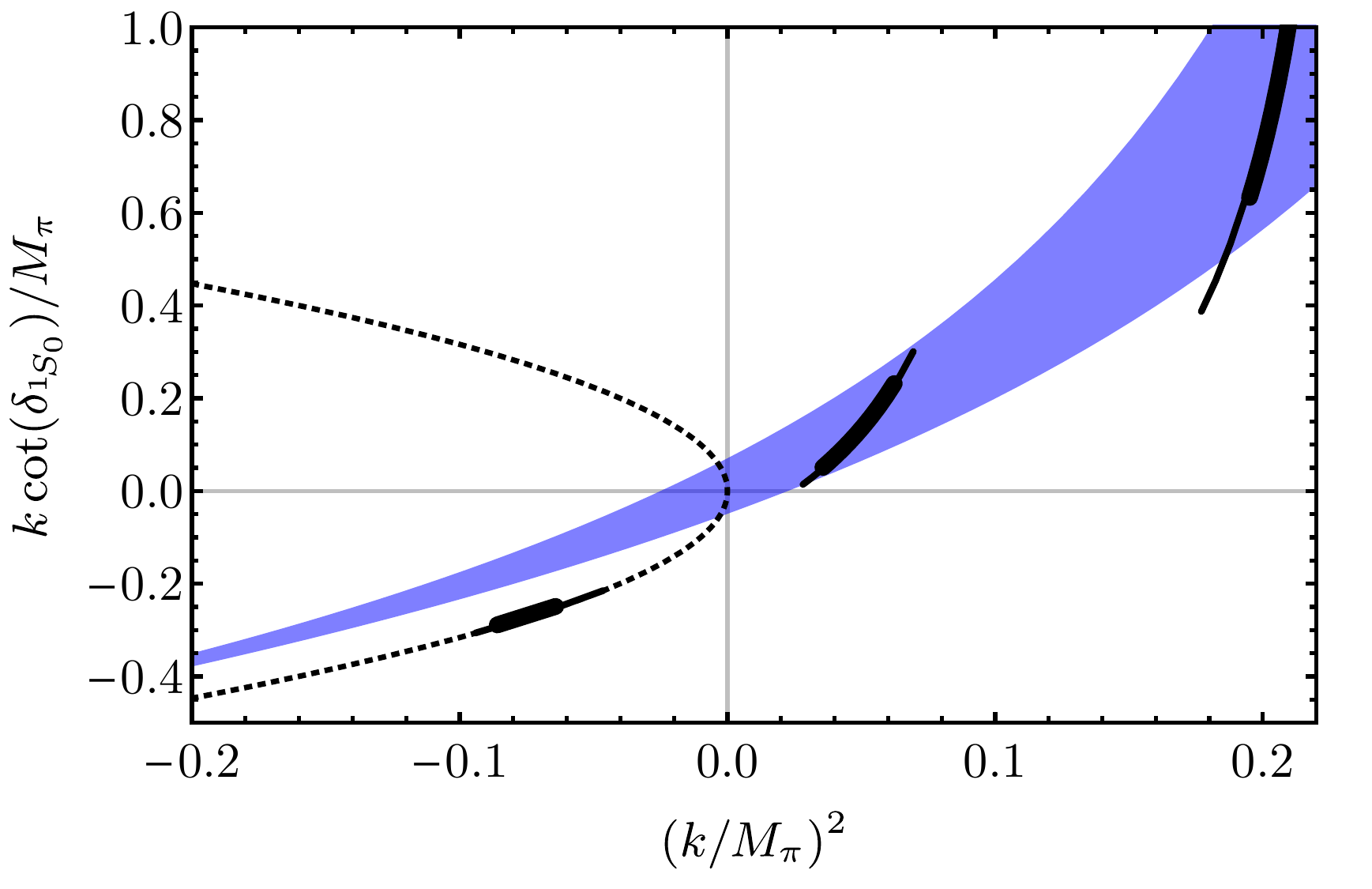}
    \caption{(Color online) Two-nucleon phase shifts (left panel) and the
      effective-range function (right panel) in the  $^1S_0$ channel
      calculated on the lattice at $M_\pi\sim
      450$~MeV~\cite{Orginos:2015aya} in
      comparison with the predictions based on the LETs at NLO (blue
      shaded bands) using
      the scattering length in Eq.~(\ref{aInvPhases}) as
      input. For remaining notation see Fig.~\ref{fig:3S1LET}. 
\label{fig:1s0_LETsa}
}
\end{figure}

As shown in Fig.~\ref{fig:1s0_LETsEB}, we arrive
at similar conclusions as in the case of the spin-triplet channel. 
While our NLO LET predictions  for  $k> 300$~MeV  are in  very good agreement with the phase
shifts calculated by the NPLQCD collaboration, there is a clear
discrepancy for the two lowest values of the momentum $k$.  In particular,
for the lowest momentum of $k\sim 100$~MeV,  the  phase shift  from the
NLO LETs is roughly a factor of two larger than that from  the
lattice-QCD analysis. Similarly to the $^3S_1$ channel, the
predictions of the LETs based on the dineutron binding energy 
are only compatible with positive values of the scattering length, see
the right panel of Fig.~\ref{fig:1s0_LETsEB}. Specifically, we
obtain  
\begin{eqnarray}\label{ERE1s0new}
{\big( M_\pi a_\text{LET, LO}^{(^1\!S_0)} \big)}^{-1}  &=&
0.244 \big({}_{-0.051}^{+0.026} \big), \quad \quad \mbox{\hspace{1.4cm}}
\!
M_\pi r_\text{LET, LO}^{(^1\!S_0)} \ =\
0.90 \big({}_{-0.06}^{+0.14} \big)\,, \nn
{\big( M_\pi a_\text{LET, NLO}^{(^1\!S_0)} \big)}^{-1}  &=&
0.175 \big({}^{+0.013}_{-0.028} \big) \big({}^{+0.024}_{-0.008} \big) 
, \quad \quad
\! M_\pi r_\text{LET, NLO}^{(^1\!S_0)} \ =\
2.86 \big({}^{+0.27}_{-0.12} \big) \big({}^{+0.27}_{-0.74} \big) 
,
\end{eqnarray}
which correspond to the following values in units of fm
\begin{eqnarray}\label{ERE1s0fm}
    a_\text{LET, LO}^{(^1\!S_0)}   &=& 
1.797\big({}^{+0.479}_{-0.171} \big)  \text{ fm}, \quad \quad \mbox{\hspace{1.4cm}}
\!
    r_\text{LET, LO}^{(^1\!S_0)}  \ =\ 
0.40\big({}^{+0.06}_{-0.03} \big)   \text{ fm}\,, \nn
    a_\text{LET, NLO}^{(^1\!S_0)}  &=& 
2.501\big({}^{+0.481}_{-0.174} \big) \big({}^{+0.123}_{-0.304} \big) 
\text{ fm},
\quad \quad
\!
    r_\text{LET, NLO}^{(^1\!S_0)} \ =\ 
1.25 \big({}^{+0.12}_{-0.05} \big) \big({}^{+0.12}_{-0.32} \big) \text{ fm}.
\end{eqnarray}
Here, the errors at LO and in the first brackets at NLO correspond to
the uncertainty in the dineutron binding energy while the ones in the
second brackets at NLO reflect the unknown $M_\pi$-dependence of
$\beta$ subject to the constraint $\delta \beta = 1$. These results
are in conflict with the NPLQCD determination based on the effective
range expansion, namely \cite{Orginos:2015aya}:
\begin{eqnarray}
\label{ERE1s0NPLQCD}
\big( M_\pi a^{(^1 \hskip -0.025in S _0)} \big)^{-1}  =  
  0.021\big({}^{+0.028}_{-0.036}\big) \big({}^{+0.032}_{-0.063} \big),  \quad \quad  M_\pi r^{(^1 \hskip
  -0.025in S _0)} \ =\
6.7 \big({}^{+1.0}_{-0.8} \big) \big({}^{+2.0}_{-1.3} \big)\,.
\end{eqnarray}
Again, we believe that the analysis performed by the NPLQCD
collaboration and based on the effective range approximation is not
self-consistent.  All arguments given in the previous section
apply to the $^1S_0$ channel too, even though our conclusions in
this case are somewhat less stringent due to the lower accuracy of the
LETs. To further elaborate on this point and to provide an assessment
of the robustness of our conclusions, we have re-done the calculations
by using the lattice phase shifts instead of the dineutron binding
energy as input. Specifically, we vary the scattering length, which is
now used as input for the LETs at NLO, in the
range consistent with the lattice-QCD phase shifts at the two lowest
energies. The resulting phase shifts, corresponding to 
 the inverse scattering
length in the range of 
\begin{equation}
\label{aInvPhases}
{\big( M_\pi a^{(^1\!S_0)} \big)}^{-1} = -0.01 \pm 0.06
\end{equation}
 are shown in the left panel of Fig.~\ref{fig:1s0_LETsa}. 
Here, we set $\delta \beta =0$, and the  
width of the band reflects the
uncertainty of the lattice-QCD phase shifts used as input. 
Notice that while the NPLQCD value of the inverse scattering length 
given in Eq.~(\ref{ERE1s0NPLQCD}) is indeed consistent with the range
of values in Eq.~(\ref{aInvPhases}), the obtained solutions correspond
to the bound (virtual) state binding energy of $B_{nn} < 0.5$~MeV ($B_{nn}^{\rm
  virtual} < 0.6$~MeV) which is in conflict with the lattice-QCD
prediction.  The apparent bound state corresponding to the leftmost intersection point
of the gray bands with the unitarity  term $ik/M_\pi=-\sqrt{-(k/M_\pi)^2}$ in the right panel of
Fig.~\ref{fig:1s0_LETsEB} is an artifact of the effective range
approximation.

Finally, it is interesting to compare our results for the
scattering length and effective range with the values obtained in
Ref.~\cite{Orginos:2015aya} within the KSW approach 
to chiral EFT \cite{Kaplan:1998we,Kaplan:1998tg}, namely
\begin{eqnarray}
\label{ERE1s0KSW}
a_{\rm KSW, \, NLO}^{(^1S_0)} &=& 2.62(07)(16)\mbox{ fm}, \quad \quad
                                   r_{\rm KSW, \, NLO}^{(^1S_0)} =
                                   1.320(18)(38)\mbox{ fm}, \nn
a_{\rm KSW, \, NNLO}^{(^1S_0)} &=& 2.99(07)(15)\mbox{ fm}, \quad \quad
                                   r_{\rm KSW, \, NNLO}^{(^1S_0)} =
                                   1.611(42)(83)\mbox{ fm}.
\end{eqnarray}
Notice that the effective range vanishes at LO in the KSW approach, and
the number of independent parameters fitted to lattice data is equal
to one, two and three at LO, NLO and next-to-next-to-leading order
(NNLO), respectively. Our NLO LET results are in excellent agreement
with the NLO KSW values and also nearly consistent with the NNLO KSW
results. 

\section{The   effective range  at unphysical pion masses}
\label{sec:rangempi}
\begin{figure}[tb]
\vspace*{-0.cm}
\includegraphics[width=0.95\textwidth,keepaspectratio,angle=0,clip]{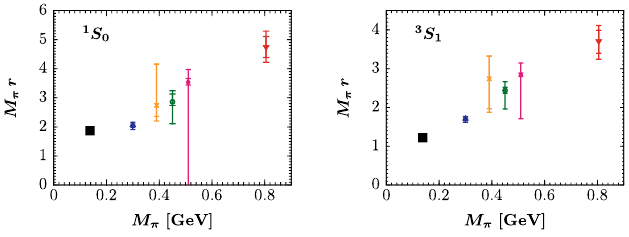}
    \caption{
\label{fig:eff_rad_mpi} (Color online) 
Nucleon-nucleon effective range      ($M_\pi r$)  in the $^1$S$_0$ (left panel) and $^3$S$_1$
      (right panel) partial waves     predicted based on the next-to-leading
      order LETs   using the bound state energies calculated on the lattice as input. 
       Blue open triangles, orange crosses, green open
      circles and purple solid circles show  $M_\pi r$     for   the binding energies       at  $M_\pi \simeq 300$~MeV \cite{Yamazaki:2015asa}, $M_\pi \simeq
      390$~MeV \cite{Beane:2011iw}, $M_\pi \simeq 450$~MeV
      \cite{Orginos:2015aya} and $M_\pi \simeq 510$~MeV
      \cite{Yamazaki:2012hi},   respectively.    
The uncertainty of our  results for  $M_\pi r$     is twofold:  the smaller error bars reflect the uncertainty of
the lattice results for the binding energies used as input      while  larger ones correspond to 
  the theoretical uncertainty of the LETs  estimated by setting $\delta \beta = 1$  and the uncertainty of
the lattice results added in quadrature. 
Red solid triangles correspond to the
      NPLQCD results for $M_\pi r$    at $M_\pi \simeq 800$~MeV \cite{Beane:2012vq}. 
The black squares show 
      the empirical values of the effective range at the
      physical pion mass \cite{deSwart:1995ui,PavonValderrama:2005ku}.
}
\end{figure}

 As already discussed,   
Ref.~\cite{Beane:2013br}  conjectured  that      the effective range   calculated on  lattice    at $M_\pi \simeq 800$~MeV and expressed in units of the pion mass
may be extrapolated to the physical point by a linear  function of $M_\pi$.   
We are now in the position to test this hypothesis by an explicit calculation based on the LETs.
Using the NN bound state energies calculated on the lattice
      at  $M_\pi \simeq 300$~MeV \cite{Yamazaki:2015asa}, $M_\pi \simeq
      390$~MeV \cite{Beane:2011iw}, $M_\pi \simeq 450$~MeV
      \cite{Orginos:2015aya} and $M_\pi \simeq 510$~MeV
      \cite{Yamazaki:2012hi},  we  employ the LETs at NLO to predict  the  values of the  effective range in the $^1$S$_0$   and $^3$S$_1$
      partial waves.  
  The results    are  visualized in Fig.~\ref{fig:eff_rad_mpi}  \footnote{The value for $M_\pi r^{(^1S_0)}$
  at $M_\pi \simeq 800$~MeV given in Ref.~\cite{Beane:2013br}, $M_\pi
   r^{(^1S_0)} = 4.61({}^{+0.29}_{-0.31})({}^{+0.24}_{-0.26})$, is somewhat different from the one plotted in their Fig.~11 and
  corresponding to the linear extrapolation specified in Eq.~(8) of
  that work. The lattice-QCD result at $M_\pi \simeq 800$~MeV shown in
  the left panel of Fig.~\ref{fig:eff_rad_mpi} is based on the
  linear extrapolation specified in Eq.~(8) of
  Ref.~\cite{Beane:2013br}.}.    
  Note that the  last  point  in both panels  at   $M_\pi \simeq 800$~MeV  represents  the result of  lattice calculations by NPLQCD~\cite{Beane:2013br},  
  while the LETs are already beyond their range of validity at such heavy pion masses. 
   As seen from the right panel  of Fig.~\ref{fig:eff_rad_mpi},  the NLO  LETs predictions for the effective range in the  $^3$S$_1$ partial wave 
   are in  very good agreement with the   linear in $M_{\pi}$ behavior of the quantity  $M_\pi r^{(^3 \hskip -0.025in S _1)}$.  Interestingly,  
  the lattice data point at   $M_\pi \simeq 800$~MeV  is    consistent  with   the NLO LET results linearly extrapolated  to  higher  pion masses.  
 The results from the LETs for the  $^1$S$_0$ partial wave,  although less conclusive 
due to larger uncertainties,  are also generally consistent with  the linear  in  $M_\pi$ behavior of $M_\pi r^{(^1 \hskip -0.025in S _0)}$. 
We note at this point that the  lattice data   at $M_\pi \simeq 800$~MeV  
were obtained in Ref.~\cite{Beane:2012vq} by using the effective range approximation.
The same procedure was employed by the NPLQCD to extract the scattering length and effective range at $M_\pi \simeq 450$~MeV and criticized in this work.
It is conceivable  that the data at  $M_\pi \simeq 800$~MeV
might also suffer  from underestimated  systematic uncertainties.

\section{Summary}
\label{sec:Summary}

In this paper, we have employed the low-energy theorems for NN scattering,
which have been generalized in Ref.~\cite{Baru:2015ira} to the case of unphysical
pion masses, to analyze the recent lattice-QCD results at $M_\pi
\simeq 450$~MeV reported by the
NPLQCD collaboration \cite{Orginos:2015aya}. The pertinent results of our work can be
summarized as follows. 
\begin{itemize}
\item
We have used the LETs along with the lattice-QCD results for the
deuteron and dineutron binding energies in order to extract the energy
behavior of the NN phase shifts in the $^3S_1$ and $^1S_0$ partial waves and the mixing
angle $\bar\epsilon_1$ at $M_\pi \simeq 450$~MeV. Our LO and NLO
calculations suggest a good (fair) convergence of our 
theoretical approach in the spin-triplet (spin-singlet) channel. In
both channels, the resulting phase shifts are in  good agreement with
the lattice-QCD results of Ref.~\cite{Orginos:2015aya} for momenta of $k > 300$~MeV,
but are inconsistent with the lattice-QCD predictions at lower
energies. 
\item
We have used the LETs to extract the values of the scattering length and effective
range in the $^3S_1$ and $^1S_0$ partial waves from the bound state
energies obtained on the lattice. The extracted value of $M_\pi r^{({}^3S_1)}$
is in excellent agreement with the linear in $M_\pi$ behavior of this
quantity conjectured in Ref.~\cite{Beane:2013br}. On the other hand,  our results
are in  strong disagreement with the values obtained by the NPLQCD
collaboration from fits to the lattice-QCD data based on the effective
range approximation. We 
have argued that the very large values for the effective range found
in Ref.~\cite{Orginos:2015aya} make the effective range approximation invalid in the
energy region corresponding to the lattice data.  
\item
Our results for phase shifts, scattering lengths and effective ranges
agree reasonably well with those obtained in Ref.~\cite{Orginos:2015aya}
by analyzing lattice-QCD data within the various EFT approaches. 
\end{itemize}
Given considerable evidence of a bound dineutron and a stronger bound
deuteron at heavy pion masses
\cite{Beane:2011iw,Yamazaki:2015asa,Yamazaki:2012hi,Orginos:2015aya},
our findings indicate that the 
lattice-QCD calculations of the NN phase shifts of Ref.~\cite{Orginos:2015aya} using the extended
L\"uscher approach may possibly suffer from
underestimated systematic errors at the lowest considered energies. 

In addition,     using    lattice results for the binding energies of  the deuteron and dineutron  at various pion masses 
as input  \cite{Beane:2011iw,Yamazaki:2015asa,Yamazaki:2012hi,Orginos:2015aya},  we  demonstrate that  the  effective range 
expressed in units of the pion mass behaves as   a linear  function of  $M_{\pi}$.

Our work demonstrates that the LETs provide a useful tool to analyze 
lattice QCD results for the NN system by allowing one to extract the
scattering phase shifts from the calculated bound state energies
and/or test consistency of lattice calculations if 
several observables are computed.

\vspace*{-0.5cm}
\acknowledgments
\vspace*{-0.2cm}
We would like to thank Jambul Gegelia for sharing his insights into
the topic discussed here and for a careful reading of the manuscript. 
We are also grateful to the organizers of the YIPQS Long-term and
Nishinomiya-Yukawa Memorial International workshop 
on ``Computational Advances in Nuclear and Hadron Physics''  at Yukawa Institute for Theoretical
Physics, Kyoto University and to the organizers of the INT-16-1 program 
``Nuclear physics from lattice QCD'' at INT Seattle, where a part of this work was carried out. 
Work supported in part by the ERC (project
259218 NuclearEFT), the BMBF (Verbundprojekt 05P2015 - NUSTAR R\&D) 
and  the DFG (grant GZ: BA 5443/1-1 AOBJ:  616443).

\end{document}